\documentclass[twocolumn,aps,prl,superscriptaddress]{revtex4-1}
\usepackage[utf8]{inputenc}
\usepackage{color}
\usepackage{bm}
\usepackage{amsmath}
\usepackage{amssymb}
\usepackage{graphicx}
\usepackage[bookmarks=false,
 breaklinks=true,pdfborder={0 0 0},pdfborderstyle={},backref=false,colorlinks=true]
 {hyperref}

\makeatletter
\usepackage{CJKutf8}
\usepackage{physics}

\makeatother

\begin{document}
\title{Quantum Thermodynamic Integrability for Canonical and Noncanonical
Statistics}
\author{R.X. Zhai (\begin{CJK}{UTF8}{gbsn}翟若迅\end{CJK})}
\affiliation{Graduate School of China Academy of Engineering Physics, Beijing 100193,
China}
\author{C.P. Sun (\begin{CJK}{UTF8}{gbsn}孙昌璞\end{CJK})}
\email{suncp@gscaep.ac.cn}

\affiliation{Graduate School of China Academy of Engineering Physics, Beijing 100193,
China}
\affiliation{School of Physics, Peking University, Beijing 100871, China}
\date{\today}
\begin{abstract}
We extend the Carathéodory principle of the second law to quantum
thermodynamics, where energy levels depend on macroscopic variables
such as volume and magnetic field. This extension introduces the concept
of quantum thermodynamic integrability (QTI), providing an alternative
foundation for statistical mechanics. QTI is characterized by the
path independence of work and heat within the thermodynamic manifold,
locally described by energy levels and specific thermodynamic parameters.
Within this framework, temperature naturally emerges as an integrating
factor, enabling the derivation of both canonical and noncanonical
states from the entropy integrable equations (EIE) based on QTI. Notably,
noncanonical states, which become particularly significant outside
the thermodynamic limit, reveal the existence of informational correlations
in finite-size thermodynamic systems.
\end{abstract}
\maketitle
\textit{Introduction--} Statistical mechanics provides the microscopic
foundation of thermodynamics, bridging the gap between the macroscopic
properties of systems and their underlying microscopic behaviors \citep{Schroedinger1952,Hatsopoulos1965,Pauli1973,Gallavotti1999,Ma2004,Landau2013}.
It aims to elucidate thermodynamic phenomena through the statistical
behavior of a large number of particles. Despite its profound significance,
the foundational principles of statistical mechanics remain subjects
of ongoing debate \citep{Khinchin1949,Jaynes1957,Jaynes1957a,Goldstein2006,Dong2007,Landau2013,Gogolin2016}.
Key issues include the use of ensembles versus time averages \citep{Khinchin1949,Landau2013},
the equal \textit{a priori} probability assumption \citep{Khinchin1949,Landau2013},
the maximum entropy principle \citep{Jaynes1957,Jaynes1957a}, and
canonical typicality based on quantum entanglement \citep{Goldstein2006,Popescu2006,Dong2007,Gogolin2016}.
However, statistical mechanics relies on these foundational assumptions
to construct statistical distributions, which lack direct experimental
validation. This raises a fundamental inverse question: Can the statistical
distribution of systems be derived directly from the fundamental principles
of thermodynamics, which have been solidly supported by experimental
evidence since the era of Watt’s steam engine?

The fundamental principles of thermodynamics were even thought to
emerge from the concept of thermodynamic integrability, as originally
proposed by Carathéodory \citep{Caratheodory1909,Born1921,LANDSBERG1964,Pauli1973,Pogliani2000,Ma2004,Zoric2011,Ma2021}.
Unlike the conventional formulation of the second law of thermodynamics,
which relies on the cycle operation of heat engines, Carathéodory’s
axiomatization is grounded in classical thermodynamic integrability
established by adiabatic inaccessibility. Such inaccessibility necessitates
the introduction of an additional thermodynamic variable to fully
describe the thermodynamic state of a system \citep{Caratheodory1909,Born1921,Pauli1973,Ma2004,Zoric2011}
and is characterized by the path independence of work, heat and entropy
in spaces defined by macroscopic mechanical variables, such as volume
and magnetic field. In the realm of quantum thermodynamics, internal
energy, work, and heat are all described in terms of discrete energy
levels and their distributions \citep{Schroedinger1952,Quan2005,Quan2007}.
To extend Carathéodory’s approach to quantum thermodynamics, we introduce
additional thermodynamic quantities that complement the mechanical
variables controlling the energy levels.

In this Letter, we align with the concept of quantum thermodynamic
integrability (QTI) as an extension of the Carathéodory principle.
As a core result, the entropy integrable equations (EIE) derived from
QTI provide a comprehensive description of various distributions in
statistical mechanics, encompassing both canonical and noncanonical
states. At equilibrium with detailed balance, the ratio of the probabilities
of two energy levels is independent of any other levels, and the canonical
states with Boltzmann distribution are uniquely determined by QTI.
We also show that, as more general solutions of the EIEs, noncanonical
states arise in finite systems outside the thermodynamic limit, which
are of particular interest. These noncanonical states imply the presence
of informational correlations and often represent statistical distributions
characterized by a superposition of different temperatures. A notable
example of a noncanonical state is Hawking radiation emitted from
small black holes, which is highly relevant to the information loss
paradox \citep{Hawking1975,Hawking1976,Parikh2000,Zhang2009,Cai2016,Dong2014}.

\textit{Quantum thermodynamic integrability --} Consider an $N$-level
quantum system with a Hamiltonian $H(\bm{\lambda})=K+V(\bm{\lambda})$,
which depends on the macroscopic mechanical variables $\bm{\lambda}=(\lambda_{1},\lambda_{2},\ldots,\lambda_{M})\in\mathbb{R}^{n}$,
such as the volume and magnetic field. Here, $K$ represents the kinetic
energy, and $V(\bm{\lambda})$ denotes the potential. The eigenstates
$|n(\bm{\lambda})\rangle$ and energy levels $E_{n}=E_{n}(\bm{\lambda})$
are controlled by $\bm{\lambda}$ (sometimes subject to boundary conditions).
These energy levels form the energy submanifold (ESM): $\{\bm{E}=(E_{1},E_{2},\ldots,E_{N})\}$.
For example, the energy levels of a free particle trapped in a one-dimensional
box are given by $E_{n}=\pi^{2}\hbar^{2}n^{2}/2mL^{2}$, where $L$
is the width of the box.

In the eigenstates $|n(\bm{\lambda})\rangle$, defined by $H(\bm{\lambda})|n(\bm{\lambda})\rangle=E_{n}(\bm{\lambda})|n(\bm{\lambda})\rangle$,
the probability distribution $P_{n}=P_{n}(\bm{E})$ forms a vector
$\bm{P}\equiv(P_{1},P_{2},\ldots,P_{N})$ that depends on $\bm{E}$.
When the potential well is adjusted slowly by $\bm{\lambda}$, the
system adiabatically remains in the state $|n(\bm{\lambda})\rangle$,
and its work changes as 
\begin{equation}
\delta W_{n}=-\langle n|\partial V/\partial\bm{\lambda}|n\rangle\cdot\text{d}\bm{\lambda}=-(\partial E_{n}/\partial\bm{\lambda})\cdot\mathrm{d}\bm{\lambda},
\end{equation}
according to the Feynman-Hellmann theorem \citep{Hellmann1933,Feynman1939}.
Therefore, the change in the average work reads $\delta W=\bm{X}\cdot\mathrm{d}\bm{\lambda}$
(like $\delta W=P\mathrm{d}V$ for ideal gases), where $\bm{X}\equiv\sum_{n}P_{n}\partial E_{n}/\partial\bm{\lambda}$
is the generalized force corresponding to $\bm{\lambda}$. In a general
situation, the variation of work is rewritten as $\delta W=\mathrm{tr}(\rho\mathrm{d}H)$
in terms of the density matrix $\rho$. Thus, a change in the internal
energy $U=\mathrm{tr}(\rho H)$ can be separated into the work $\delta W$
and the heat $\delta Q=\mathrm{tr}(H\mathrm{d}\rho)$, accordingly.

\begin{figure}
\begin{centering}
\includegraphics{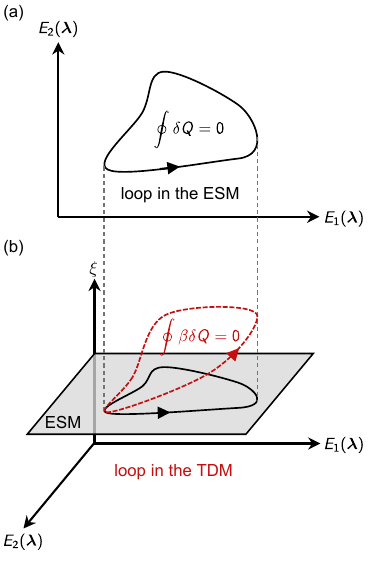}
\par\end{centering}
\caption{\label{fig:path_independent_sketch} Extending from ESM to the TDM
shown with two-level system. (a) In the energy submanifold, the vanishing
work integration over a close path implies the QTI. (b) In the thermodynamics
manifold where the ESM is its submanifold the vanishing entropy integration
over a close path implies the QTI.}
\end{figure}

According to classical Carathéodory principle, \textit{``in any arbitrary
neighborhood of an arbitrarily given initial point there is a state
that cannot be arbitrarily approximated by adiabatic changes of state''}
\citep{Caratheodory1909,Born1921,Pauli1973,Ma2004,Zoric2011}. Such
adiabatic inaccessibility necessitates that solely mechanical variables
are insufficient to fully describe the thermodynamic state; an additional
thermodynamic variable must be introduced. Thus, to generalize the
Carathéodory principle to the quantum realm, we assume that the density
matrix for the equilibrium state depends not only on $\bm{\lambda}$,
but also on some thermodynamic variables $\xi$ to be considered,
namely, $\rho=\rho(\bm{\lambda},\xi)$. In the thermodynamic manifold
(TDM): $\{[\bm{E}(\bm{\lambda}),\xi]\}$ parameterized by $\{(\bm{\lambda},\xi)\}$
(see FIG. \ref{fig:path_independent_sketch}), the heat reads as 
\[
\delta Q=\mathrm{tr}(H\partial_{\mu}\rho)\mathrm{d}\lambda_{\mu}+\mathrm{tr}(H\partial_{\xi}\rho)\mathrm{d}\xi.
\]
Here, $\partial_{\mu}\equiv\partial/\partial\lambda_{\mu}$, $\partial_{\xi}\equiv\partial/\partial\xi$,
and a repeated index in a term implies summation. Obviously, the Carathéodory
principle can work well in the parametrized TDM : $\{(\bm{\lambda},\xi)\}$,
establishing the so-called QTI.

\textit{Entropy integrable equations for statistical distribution}
-- The QTI requires the existence of an integrating factor $\beta=\beta(\bm{\lambda},\xi)$
in TDM such that the thermodynamic entropy change $\delta S\equiv\beta\delta Q$
is a total differential, guaranteed by $\mathrm{d}(\delta S)=0$.
This leads to an equation in two-form describing the integrability
of entropy. We consider a special solution $\beta=\xi$, which is
the simplest integrating factor. Here, $\beta=1/k_{B}T$ can be interpreted
as the inverse temperature. Consequently, the requirement of QTI is
expressed as
\begin{align}
\mathrm{tr}\left(\partial_{\mu}H\partial_{\nu}\rho-\partial_{\nu}H\partial_{\mu}\rho\right) & =0,\label{eq:EIE_rho_H_01}\\
\mathrm{tr}\left(H\partial_{\mu}\rho-\beta\partial_{\mu}H\partial_{\beta}\rho\right) & =0,\label{eq:EIE_rho_H_02}
\end{align}
where $\partial_{\beta}\equiv\partial/\partial\beta$. It is noteworthy
that the QTI for the composed system guarantees the universality of
the integrating factor $\beta$. For example, consider a system separated
into two parts, $1$ and $2$. Let the integrating factor of the two
parts be denoted as $\beta_{1}$ and $\beta_{2}$. Consequently, the
thermodynamic entropy for subsystem $i=1,2$ is given by $\mathrm{d}S_{i}=\beta_{i}\delta Q_{i}$.
The additivity of heat $\delta Q=\delta Q_{1}+\delta Q_{2}$ ensures
the additivity of entropy $\mathrm{d}S=\mathrm{d}S_{1}+\mathrm{d}S_{2}$,
when $\beta_{1}=\beta_{2}=\beta$. In other words, the integrating
factor $\beta$ is universal across all systems in equilibrium.

Typically, with the diagonal representation of density matrix, $\rho=\sum_{n}P_{n}|n\rangle\langle n|$,
the probability distribution $P_{n}=P_{n}[\bm{E}(\bm{\lambda}),\beta]$
depends on $\bm{\lambda}$ through $\bm{E}(\bm{\lambda})$. Thus,
Eqs. \eqref{eq:EIE_rho_H_01} and \eqref{eq:EIE_rho_H_02} are reduced
to

\begin{align}
\sum_{m<n}\left(\partial_{m}P_{n}-\partial_{n}P_{m}\right)\{E_{m},E_{n}\}_{\mu\nu} & =0,\\
\left(\partial_{\mu}\bm{E}\right)\cdot\left[(\bm{\bm{E}\cdot\nabla})-\beta\partial_{\beta}\right]\bm{P} & =0,
\end{align}
where $\partial_{n}\equiv\partial/\partial E_{n}$, $\{A,B\}_{\mu\nu}\equiv(\partial_{\mu}A\partial_{\nu}B-\partial_{\nu}A\partial_{\mu}B)$,
and $\bm{\nabla}\equiv\left(\partial_{1},\partial_{2},\cdots,\partial_{N}\right)$.

It is observed that the QTI is adequately ensured by the following
equations:

\begin{align}
\partial_{m}P_{n}-\partial_{n}P_{m} & =0,\label{eq:EIE_prob_E}\\
\left[(\bm{\bm{E}\cdot\nabla})-\beta\partial_{\beta}\right]\bm{P} & =0.\label{eq:EIE_prob_beta}
\end{align}
Equations. \eqref{eq:EIE_prob_E} and \eqref{eq:EIE_prob_beta}, referred
to as the EIEs, represent our core results. As sketched with two energy
levels in FIG. \ref{fig:path_independent_sketch}, the two EIEs represent
the QTI within the ESM {[}FIG. \ref{fig:path_independent_sketch}(a){]}
and TDM {[}FIG. \ref{fig:path_independent_sketch}(b){]}, respectively.

The universal solution $\bm{P}=\bm{\nabla}F(\beta,\bm{E})$ to Eq.
\eqref{eq:EIE_prob_E} can be expressed in terms of a generating function
$F$. The normalization of the probability distribution requires $e\cdot\bm{\nabla}F=1$,
for $\bm{e}=(1,1,\ldots,1)$. The function $F(\beta,\bm{E})$ can
be understood as the generalized free energy, and Eq. \eqref{eq:EIE_prob_beta}
is reduced to
\begin{equation}
\bm{E}\cdot\bm{\nabla}F=\partial_{\beta}(\beta F)=F+\beta\partial_{\beta}F.\label{eq:EIE_Free_Energy}
\end{equation}
For a detailed derivation of the EIEs, please refer to Sec. I in Supplementary
Materials \citep{SM_This}. Here, the normalization condition guarantees
that the distribution is independent of the reference point of energy.
Specifically, for a uniform shift in the energy levels $\bm{E}\rightarrow\bm{E}+\delta\bm{e}$,
we obtain the displacement symmetry
\[
P_{n}(\bm{E}+\delta\bm{e})\approx P_{n}(\bm{E})+\delta\bm{e}\cdot\nabla P_{n}=P_{n}(\bm{E}),
\]
since $\bm{e}\cdot\nabla P_{n}=\partial_{n}(\bm{e}\cdot\nabla F)=0$.

\textit{Statistical distribution from QTI }-- To obtain the distribution
vector $\bm{P}$ from the EIEs based on QTI, we consider the symmetry
of $F=F(E_{1},\ldots,E_{i},\ldots,E_{j},\ldots)$. The internal energy
$U=\sum_{n}E_{n}P_{n}$ possesses the exchanging symmetry under the
operation: $(E_{i}\leftrightarrow E_{j})$, $(P_{i}\leftrightarrow P_{j})$,
which guarantees the invariance of statistical distributions $(P_{1},P_{2},\ldots,P_{N})$.
Up to the same scaling $\beta$ (with Eq. \eqref{eq:EIE_Free_Energy}),
the EIE has a symmetric solution
\begin{equation}
F=\left(1/\beta\right)\Psi(\beta\bm{E})=\left(1/\beta\right)\Psi(\beta E_{1},\beta E_{2},\ldots,\beta E_{N}),
\end{equation}
where $\Psi$ is a symmetric function of $N$ variables.

It should be noted that the $E_{n}$ appears in $F$ only in the form
of its product with $\beta$. The QTI reflects the fact that the thermodynamic
entropy $S$ is a state function. We can verify that $\mathrm{d}S=\beta\delta Q$,
as a total differential, satisfies $S=\beta(U-F)$, which refers to
the second law of thermodynamics. Following Carathéodory's arguments,
it can be proved within our framework that the entropy of an arbitrary
adiabatic process never decreases \citep{Caratheodory1909,Born1921,Pogliani2000}.

We observe that the statistical distribution $P_{n}(\bm{E})=\partial_{n}\left(\Psi(\beta\bm{E})/\beta\right)$
usually depends on all energy levels. When the interaction between
the system and the reservoir is weak, the transition rate $W_{m\rightarrow n}$
between two states $m\rightarrow n$ does not depend on any intermediate
states $l\neq m,n$. Thus, for equilibrium states with the principle
of detailed balance \citep{Breuer2007}, the relative probability
$P_{n}(\bm{E})/P_{m}(\bm{E})$ also does not depends on any other
level $E_{l}$ ($l\neq m,n$). In this sense, we can prove (see Sec.
II B in Supplementary Materials \citep{SM_This}) that the symmetric
function $\Psi(x_{1},x_{2},\ldots,x_{N})$ reduces to $\Psi=\Psi[\sum_{n=1}^{N}w(x_{n})]$,
which corresponds to
\[
P_{n}=\frac{w'(E_{n})}{\sum_{m}w'(E_{m})}.
\]
The EIE further determines the concrete form of $w$ as $w(E_{n})=\exp(-\beta E_{n})$,
resulting in the Boltzmann distribution
\[
P_{n}=\frac{1}{Z}\mathrm{e}^{-\beta E_{n}},
\]
where $Z=\sum_{n}\exp(-\beta E_{n})$ is the partition function with
respect to the free energy function $F=-k_{B}T\ln Z$. Here, we choose
$\beta$ to be positive, which can be sufficiently determined with
one experiment \citep{Caratheodory1909,Born1921}. For more discussion
of this derivation, please see Sec. II in Supplementary Material \citep{SM_This}.

\textit{Noncanonical states from EIE} - Noncanonical statistics are
typically shaped by interactions between the system and reservoir
\citep{Xu2014,Richens2018,Chung2019}. When the interaction becomes
stronger, the backaction of the system on the reservoir becomes non-negligible.
As a result, the transition between $m$ and $n$ is influenced by
some intermediate states $E_{l}$ ($l\neq m,n$) through second- or
higher-order transitions, violating the principle of detailed balance.
Such effect is particularly evident when the system is small and outside
the thermodynamic limit. The temperature of such nonequilibrium steady
states is not well defined. Instead, we can effectively describe them
by assuming the inverse temperature distributes around a given integrating
factor $\beta$ with a standard deviation $\Delta\beta$ \citep{Beck2003,Falcioni2011,Fei2024}.

Specifically, a linear superposition of free energy $F^{(\mathrm{nc})}(\beta,\bm{E})=\int\mathrm{d}\beta'f(\beta,\beta')F(\beta',\boldsymbol{E})$
over varying temperatures results in a noncanonical solution to the
EIE 
\begin{equation}
P_{n}^{(\mathrm{nc})}(\beta)=\int\mathrm{d}\beta'f(\beta,\beta')P_{n}(\beta').\label{eq:non_can_dist}
\end{equation}
Here, the coefficient $f(\beta,\beta')$ satisfies the equation $\partial_{\beta}[\beta f(\beta,\beta')]+\beta'\partial_{\beta'}f(\beta,\beta')=0$
(see Sec. III in Supplementary Material \citep{SM_This} for details).
Without loss of generality, we can rescale $\beta$ so that it represents
the average inverse temperature $\beta=\int\mathrm{d}\beta'f(\beta,\beta')\beta'$.

When the standard deviation $\Delta\beta$ of $\beta$ is sufficiently
small, the noncanonical states (\ref{eq:non_can_dist}) are expanded
as
\begin{equation}
P_{n}^{(\mathrm{nc})}(\beta,\Delta\beta)\approx\frac{1}{Z_{\mathrm{eff}}}\exp[-\beta_{\mathrm{eff}}E_{n}+(\Delta\beta^{2}/2)E_{n}^{2}],\label{eq:dist_expand}
\end{equation}
where $\beta_{\mathrm{eff}}\equiv\beta+U\Delta\beta^{2}$, $Z_{\mathrm{eff}}\equiv Z\mathrm{e}^{-\Delta\beta^{2}(U^{2}-\left\langle \Delta E^{2}\right\rangle )/2}$,
and $\left\langle \Delta E^{k}\right\rangle $ is the $k$th central
moment of energy for the canonical states. With a small deviation,
the detailed shape of the function $f(\beta,\beta')$ becomes inessential,
where all the noncanonical thermodynamic quantities are expressed
as functions of the standard deviation $\Delta\beta$ and equilibrium
quantities (see Sec. IV in Supplementary Materials \citep{SM_This}).
Therefore, without loss of generality, the coefficient $f(\beta',\beta)$
can be chosen as a Gaussian distribution $f(\beta,\beta')\sim\mathcal{N}(\beta^{2},\Delta\beta^{2})$
in the following calculations. It follows from the EIE that the relative
deviation of inverse temperature $\sigma\equiv\Delta\beta/\beta$
for such states remains constant.

\begin{figure}
\centering{}\includegraphics{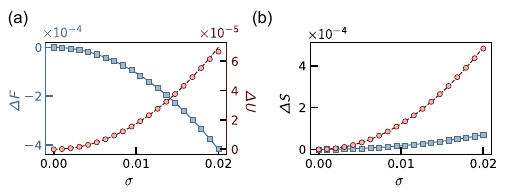} \caption{\label{fig:compare_can_noncan_distdiff} Difference in the canonical
and noncanonical states for three-level system where $\Delta_{1}=1,\Delta_{2}=2$.
The markers are obtained with numerical calculation, and the curves
depict the corresponding leading terms. (a) Difference in internal
energy $\Delta U$ (red circles and curve) and generalized free energy
$\Delta F$ (blue squares and curve). (b) Difference in information
entropy (blue squares and curve) and thermodynamic entropy (red circles
and curve).}
\end{figure}

To quantify the difference between the canonical and noncanonical
states, we investigate an example of three-level systems with $E_{0}=0$,
$E_{1}=\Delta_{1}$ and $E_{2}=\Delta_{2}$. In FIG. \ref{fig:compare_can_noncan_distdiff},
the thermodynamic functions of noncanonical states are calculated
to compare with those of canonical states, which are evaluated both
numerically and analytically as functions of the relative deviation
$\sigma$. Respectively, FIG \ref{fig:compare_can_noncan_distdiff}(a)
illustrates the difference in the internal energy $\Delta U\equiv U^{(\mathrm{nc})}-U$
(red circles and curve) and the generalized free energy $\Delta F\equiv F^{(\mathrm{nc})}-F$
(blue squares and curve), and FIG \ref{fig:compare_can_noncan_distdiff}(b)
shows the difference in information entropy (blue squares and curve)
and thermodynamic entropy (red circles and curve) for the noncanonical
distributions. In the figure, all markers are evaluated numerically,
and the curves represent the leading order shown in Sec. IV in Supplementary
Material \citep{SM_This}. Here we have chosen $\Delta_{1}=1$, $\Delta_{2}=2$,
and $\beta=1$.

\begin{figure}[!htbp]
\centering{}\includegraphics{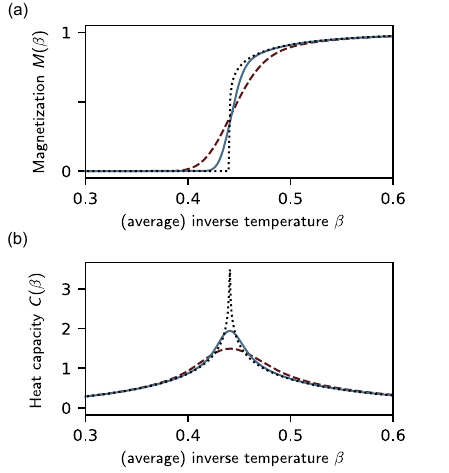} \caption{\label{fig:non-can_Ising} Thermodynamic functions of canonical and
noncanonical states of the two-dimensional Ising model. (a) Magnetization
as a function of inverse temperature $\beta$ (b) Heat capacity as
a function of inverse temperature $\beta$. The black dotted curves
represent the canonical states. The blue solid curves and the red
dashed curves represent the noncanonical states with $\Delta\beta/\beta=0.02$
and $\Delta\beta/\beta=0.05$, respectively.}
\end{figure}

Systems in the noncanonical states usually show observable effects,
especially for those with phase transitions. For canonical states,
thermodynamic functions of a phase transition system exhibit discontinuity
at the critical point, and such discontinuity is smoothed for noncanonical
states. We take the two dimensional Ising model as an example. In
canonical states, the system exhibits a phase transition at the critical
temperature $\beta_{c}=\ln(1+\sqrt{2})/2J$, where $J$ is the amplitude
of the exchange interaction between spins. Near the critical point,
the macroscopic magnetization $M$ is zero for $t<0$ and increases
sharply as $M\propto t^{1/8}$ for $t>0$, and the heat capacity $C(\beta)$
diverges as $\ln|t|$ \citep{Onsager1944,Yang1952}. Here $t\equiv(\beta-\beta_{c})/\beta$
is the relative distance to the critical point. Such critical behaviors
for canonical states are illustrated in FIG. \ref{fig:non-can_Ising}
with black dotted curves.

For the noncanonical states, the thermodynamic functions near $\beta_{c}$
are dependent on $t$ and $\sigma$. We illustrate in FIG. \ref{fig:non-can_Ising}
the noncanonical magnetization and heat capacity, demonstrating that
the critical behaviors are smoothed. The data for $\sigma=0.02$ and
$\sigma=0.05$ are shown with blue solid curves and red dashed curves,
respectively. Interestingly, the macroscopic magnetization of the
noncanonical states near $\beta_{c}$ scales as

\[
M(t,\mu\sigma)=\mu^{1/8}M(t/\mu,\sigma).
\]
If we assume that $\sigma\sim1/L$, where $L$ is the characteristic
length of the system, such noncanonical states behave the same scaling
with the finite-sized systems outside the thermodynamic limit \citep{Fisher1967,Ferdinand1969,Binder1972,Privman1984}.
The calculations related to FIG. \ref{fig:non-can_Ising} and the
finite-size scaling can be found in Sec V in Supplementary Material
\citep{SM_This}.

\begin{figure}[!htbp]
\centering{}\includegraphics{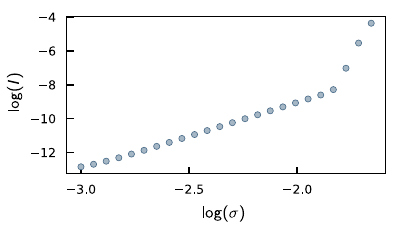} \caption{\label{fig:mutal_info}\textit{ }\textit{\emph{Mutual information
between two-three level systems as a function of the relative deviation
of inverse temperature.}}}
\end{figure}

\textit{Correlation from noncanonical states --} The noncanonical
distributions also reveal the information correlation between two
events associated with $E_{m}$ and $E_{n}$. We consider a two-component
system without interaction between the components, both of which have
the same energy levels $E_{n}$. Unlike the canonical distribution,
the joint distribution of the two components deviates from being a
simple product of the distributions of each component, as discussed
in Sec. VI in Supplementary Material \citep{SM_This}. This is mathematically
represented by the inequality $P_{mn}^{(\mathrm{nc})}(\beta)\neq P_{m}^{(\mathrm{nc})}(\beta)P_{n}^{(\mathrm{nc})}(\beta)$.
Thus, the information entropy $S_{\mathrm{in}}(P_{n})=-\sum_{n}P_{n}\ln P_{n}$
for the subsystems exhibits a lack of additivity due to the noncanonical
correlation, which is quantified by the mutual information $I\equiv S_{\mathrm{in}}(P_{m}^{(\mathrm{nc})})+S_{\mathrm{in}}(P_{n}^{(\mathrm{nc})})-S_{\mathrm{in}}(P_{mn}^{(\mathrm{nc})})$.
We plot the mutual information of two correlated three-level systems
as a function of the relative deviation of $\beta$ in FIG. \ref{fig:mutal_info}.

Actually, the noncanonicality of a black hole induces the information
embedded in the correlation of its radiation, which has been considered
a promising candidate for resolving the black hole information paradox.
The question arises whether the noncanonical solutions to the EIE
can describe such a noncanonical radiation field from a small Schwarzschild
black hole. For a Schwarzschild black hole with temperature $\beta=8\pi M$,
the deviation of $\beta$ is estimated as $\Delta\beta^{2}=\left|\beta^{2}/C\right|=8\pi$,
where $C=-8\pi M^{2}$ is the heat capacity of the black hole \citep{Hawking1975,Wald2001,Landau2013,Carlip2014}.
From the noncanonical solutions of EIE, the emission rate (see Sec.
VII in Supplementary Material \citep{SM_This}) is given by
\begin{equation}
\Gamma(\beta,\omega)\approx\frac{1}{4\pi^{2}}\exp[-8\pi\omega(M-\omega/2)],
\end{equation}
which is valid in cases with small deviations. This result agrees
with the previous findings from the microscopic theory for the radiation
field near a small black hole \citep{Parikh2000}. The noncanonical
state can account for the black hole information paradox \citep{Zhang2009,Cai2016}.
It is emphasized that such noncanonical solutions presented in this
letter arises from QTI. As black holes become smaller, the backaction
of radiation grows stronger, leading to a tighter correlation between
successive radiation events. This increased correlation causes the
radiation state to deviate significantly from the traditional canonical
thermal equilibrium state.

\textit{Conclusion} -- In conclusion, we have proposed an alternative
approach to reconsider the foundations of statistical mechanics based
on the QTI introduced in this Letter as a generalization of the path
independence of work and heat. We established a quantum version of
Carathéodory’s principle and derived the so-called EIEs, which can
result in both canonical and noncanonical states. By considering detailed
balance, we uniquely constructed the Boltzmann distribution from EIE
for QTI as the foundation of statistical physics. In this sense, QTI
not only reinforces the principles of statistical physics based on
the experiment-driven principles of thermodynamics but also provides
a new perspective on noncanonical thermodynamics grounded in information
correlation. This work opens up new avenues for addressing long-standing
problems in statistical thermodynamics, offering a more comprehensive
framework for analyzing complex thermodynamic behaviors in small systems.

\textit{Acknowledgment} -- We are grateful to H. Dong, H. T. Quan,
and Y. H. Ma for their helpful discussions. This work has been supported
by the National Natural Science Foundation of China (NSFC) (Grant
No. 12088101) and NSAF No.U2330401.

\bibliography{refs}

\begin{thebibliography}{45}%
\makeatletter
\providecommand \@ifxundefined [1]{%
 \@ifx{#1\undefined}
}%
\providecommand \@ifnum [1]{%
 \ifnum #1\expandafter \@firstoftwo
 \else \expandafter \@secondoftwo
 \fi
}%
\providecommand \@ifx [1]{%
 \ifx #1\expandafter \@firstoftwo
 \else \expandafter \@secondoftwo
 \fi
}%
\providecommand \natexlab [1]{#1}%
\providecommand \enquote  [1]{``#1''}%
\providecommand \bibnamefont  [1]{#1}%
\providecommand \bibfnamefont [1]{#1}%
\providecommand \citenamefont [1]{#1}%
\providecommand \href@noop [0]{\@secondoftwo}%
\providecommand \href [0]{\begingroup \@sanitize@url \@href}%
\providecommand \@href[1]{\@@startlink{#1}\@@href}%
\providecommand \@@href[1]{\endgroup#1\@@endlink}%
\providecommand \@sanitize@url [0]{\catcode `\\12\catcode `\$12\catcode
  `\&12\catcode `\#12\catcode `\^12\catcode `\_12\catcode `\%12\relax}%
\providecommand \@@startlink[1]{}%
\providecommand \@@endlink[0]{}%
\providecommand \url  [0]{\begingroup\@sanitize@url \@url }%
\providecommand \@url [1]{\endgroup\@href {#1}{\urlprefix }}%
\providecommand \urlprefix  [0]{URL }%
\providecommand \Eprint [0]{\href }%
\providecommand \doibase [0]{http://dx.doi.org/}%
\providecommand \selectlanguage [0]{\@gobble}%
\providecommand \bibinfo  [0]{\@secondoftwo}%
\providecommand \bibfield  [0]{\@secondoftwo}%
\providecommand \translation [1]{[#1]}%
\providecommand \BibitemOpen [0]{}%
\providecommand \bibitemStop [0]{}%
\providecommand \bibitemNoStop [0]{.\EOS\space}%
\providecommand \EOS [0]{\spacefactor3000\relax}%
\providecommand \BibitemShut  [1]{\csname bibitem#1\endcsname}%
\let\auto@bib@innerbib\@empty
\bibitem [{\citenamefont {Schr{\"o}dinger}(1952)}]{Schroedinger1952}%
  \BibitemOpen
  \bibfield  {author} {\bibinfo {author} {\bibfnamefont {E.}~\bibnamefont
  {Schr{\"o}dinger}},\ }\href@noop {} {\emph {\bibinfo {title} {{S}tatistical
  {T}hermodynamics}}}\ (\bibinfo  {publisher} {{C}ambridge {U}niversity
  {P}ress},\ \bibinfo {address} {England},\ \bibinfo {year} {1952})\BibitemShut
  {NoStop}%
\bibitem [{\citenamefont {Hatsopoulos}\ and\ \citenamefont
  {Keenan}(1965)}]{Hatsopoulos1965}%
  \BibitemOpen
  \bibfield  {author} {\bibinfo {author} {\bibfnamefont {G.~N.}\ \bibnamefont
  {Hatsopoulos}}\ and\ \bibinfo {author} {\bibfnamefont {J.~H.}\ \bibnamefont
  {Keenan}},\ }\href@noop {} {\emph {\bibinfo {title} {Principles of general
  thermodynamics}}}\ (\bibinfo  {publisher} {Wiley},\ \bibinfo {address} {New
  York [u.a.]},\ \bibinfo {year} {1965})\BibitemShut {NoStop}%
\bibitem [{\citenamefont {Pauli}(1973)}]{Pauli1973}%
  \BibitemOpen
  \bibfield  {author} {\bibinfo {author} {\bibfnamefont {W.}~\bibnamefont
  {Pauli}},\ }\href@noop {} {\emph {\bibinfo {title} {Pauli lectures on physics
  : Thermodynamics and the Kinetic Theory of Gases}}},\ edited by\ \bibinfo
  {editor} {\bibfnamefont {W.}~\bibnamefont {Pauli}},\ Vol.~\bibinfo {volume}
  {3}\ (\bibinfo  {publisher} {MIT Press},\ \bibinfo {address} {Cambridge,
  Mass.},\ \bibinfo {year} {1973})\BibitemShut {NoStop}%
\bibitem [{\citenamefont {Gallavotti}(1999)}]{Gallavotti1999}%
  \BibitemOpen
  \bibfield  {author} {\bibinfo {author} {\bibfnamefont {G.}~\bibnamefont
  {Gallavotti}},\ }\href {\doibase 10.1007/978-3-662-03952-6} {\emph {\bibinfo
  {title} {Statistical Mechanics}}}\ (\bibinfo  {publisher} {Springer Berlin
  Heidelberg},\ \bibinfo {year} {1999})\BibitemShut {NoStop}%
\bibitem [{\citenamefont {Ma}(2004)}]{Ma2004}%
  \BibitemOpen
  \bibfield  {author} {\bibinfo {author} {\bibfnamefont {S.-K.}\ \bibnamefont
  {Ma}},\ }\href@noop {} {\emph {\bibinfo {title} {Statistical mechanics}}},\
  \bibinfo {edition} {3rd}\ ed.,\ edited by\ \bibinfo {editor} {\bibfnamefont
  {M.}~\bibnamefont {Fung}}\ (\bibinfo  {publisher} {World Scientific Publ.},\
  \bibinfo {address} {Philadelphia [u.a.]},\ \bibinfo {year} {2004})\ \bibinfo
  {note} {aus d. Chines. übers. - Literaturverz. S. 541 - 544}\BibitemShut
  {NoStop}%
\bibitem [{\citenamefont {Landau}\ and\ \citenamefont
  {Lifshitz}(2013)}]{Landau2013}%
  \BibitemOpen
  \bibfield  {author} {\bibinfo {author} {\bibfnamefont {L.~D.}\ \bibnamefont
  {Landau}}\ and\ \bibinfo {author} {\bibfnamefont {E.~M.}\ \bibnamefont
  {Lifshitz}},\ }\href@noop {} {\emph {\bibinfo {title} {{S}tatistical
  {P}hysics}}},\ Vol.~\bibinfo {volume} {5}\ (\bibinfo  {publisher}
  {Elsevier},\ \bibinfo {year} {2013})\BibitemShut {NoStop}%
\bibitem [{\citenamefont {Khinchin}(1949)}]{Khinchin1949}%
  \BibitemOpen
  \bibfield  {author} {\bibinfo {author} {\bibfnamefont {A.~I.}\ \bibnamefont
  {Khinchin}},\ }\href@noop {} {\emph {\bibinfo {title} {Mathematical
  foundations of statistical mechanics}}},\ Dover books on mathematics\
  (\bibinfo  {publisher} {Dover},\ \bibinfo {address} {New York},\ \bibinfo
  {year} {1949})\BibitemShut {NoStop}%
\bibitem [{\citenamefont {Jaynes}(1957{\natexlab{a}})}]{Jaynes1957}%
  \BibitemOpen
  \bibfield  {author} {\bibinfo {author} {\bibfnamefont {E.~T.}\ \bibnamefont
  {Jaynes}},\ }\href {\doibase 10.1103/physrev.106.620} {\bibfield  {journal}
  {\bibinfo  {journal} {Phys. Rev.}\ }\textbf {\bibinfo {volume} {106}},\
  \bibinfo {pages} {620} (\bibinfo {year} {1957}{\natexlab{a}})}\BibitemShut
  {NoStop}%
\bibitem [{\citenamefont {Jaynes}(1957{\natexlab{b}})}]{Jaynes1957a}%
  \BibitemOpen
  \bibfield  {author} {\bibinfo {author} {\bibfnamefont {E.~T.}\ \bibnamefont
  {Jaynes}},\ }\href {\doibase 10.1103/physrev.108.171} {\bibfield  {journal}
  {\bibinfo  {journal} {Phys. Rev.}\ }\textbf {\bibinfo {volume} {108}},\
  \bibinfo {pages} {171} (\bibinfo {year} {1957}{\natexlab{b}})}\BibitemShut
  {NoStop}%
\bibitem [{\citenamefont {Goldstein}\ \emph {et~al.}(2006)\citenamefont
  {Goldstein}, \citenamefont {Lebowitz}, \citenamefont {Tumulka},\ and\
  \citenamefont {Zangh\`i}}]{Goldstein2006}%
  \BibitemOpen
  \bibfield  {author} {\bibinfo {author} {\bibfnamefont {S.}~\bibnamefont
  {Goldstein}}, \bibinfo {author} {\bibfnamefont {J.~L.}\ \bibnamefont
  {Lebowitz}}, \bibinfo {author} {\bibfnamefont {R.}~\bibnamefont {Tumulka}}, \
  and\ \bibinfo {author} {\bibfnamefont {N.}~\bibnamefont {Zangh\`i}},\ }\href
  {\doibase 10.1103/physrevlett.96.050403} {\bibfield  {journal} {\bibinfo
  {journal} {Phys. Rev. Lett.}\ }\textbf {\bibinfo {volume} {96}},\ \bibinfo
  {pages} {050403} (\bibinfo {year} {2006})}\BibitemShut {NoStop}%
\bibitem [{\citenamefont {Dong}\ \emph {et~al.}(2007)\citenamefont {Dong},
  \citenamefont {Yang}, \citenamefont {Liu},\ and\ \citenamefont
  {Sun}}]{Dong2007}%
  \BibitemOpen
  \bibfield  {author} {\bibinfo {author} {\bibfnamefont {H.}~\bibnamefont
  {Dong}}, \bibinfo {author} {\bibfnamefont {S.}~\bibnamefont {Yang}}, \bibinfo
  {author} {\bibfnamefont {X.~F.}\ \bibnamefont {Liu}}, \ and\ \bibinfo
  {author} {\bibfnamefont {C.~P.}\ \bibnamefont {Sun}},\ }\href {\doibase
  10.1103/physreva.76.044104} {\bibfield  {journal} {\bibinfo  {journal} {Phys.
  Rev. A}\ }\textbf {\bibinfo {volume} {76}},\ \bibinfo {pages} {044104}
  (\bibinfo {year} {2007})}\BibitemShut {NoStop}%
\bibitem [{\citenamefont {Gogolin}\ and\ \citenamefont
  {Eisert}(2016)}]{Gogolin2016}%
  \BibitemOpen
  \bibfield  {author} {\bibinfo {author} {\bibfnamefont {C.}~\bibnamefont
  {Gogolin}}\ and\ \bibinfo {author} {\bibfnamefont {J.}~\bibnamefont
  {Eisert}},\ }\href {\doibase 10.1088/0034-4885/79/5/056001} {\bibfield
  {journal} {\bibinfo  {journal} {Rep. Prog. Phys.}\ }\textbf {\bibinfo
  {volume} {79}},\ \bibinfo {pages} {056001} (\bibinfo {year}
  {2016})}\BibitemShut {NoStop}%
\bibitem [{\citenamefont {Popescu}\ \emph {et~al.}(2006)\citenamefont
  {Popescu}, \citenamefont {Short},\ and\ \citenamefont
  {Winter}}]{Popescu2006}%
  \BibitemOpen
  \bibfield  {author} {\bibinfo {author} {\bibfnamefont {S.}~\bibnamefont
  {Popescu}}, \bibinfo {author} {\bibfnamefont {A.~J.}\ \bibnamefont {Short}},
  \ and\ \bibinfo {author} {\bibfnamefont {A.}~\bibnamefont {Winter}},\ }\href
  {\doibase 10.1038/nphys444} {\bibfield  {journal} {\bibinfo  {journal} {Nat.
  Phys.}\ }\textbf {\bibinfo {volume} {2}},\ \bibinfo {pages} {754} (\bibinfo
  {year} {2006})}\BibitemShut {NoStop}%
\bibitem [{\citenamefont {Carath\'eodory}(1909)}]{Caratheodory1909}%
  \BibitemOpen
  \bibfield  {author} {\bibinfo {author} {\bibfnamefont {C.}~\bibnamefont
  {Carath\'eodory}},\ }\href {\doibase 10.1007/bf01450409} {\bibfield
  {journal} {\bibinfo  {journal} {Math. Ann.}\ }\textbf {\bibinfo {volume}
  {67}},\ \bibinfo {pages} {355} (\bibinfo {year} {1909})}\BibitemShut
  {NoStop}%
\bibitem [{\citenamefont {Born}(1921)}]{Born1921}%
  \BibitemOpen
  \bibfield  {author} {\bibinfo {author} {\bibfnamefont {M.}~\bibnamefont
  {Born}},\ }\href@noop {} {\bibfield  {journal} {\bibinfo  {journal} {Phys.
  Zeit.}\ }\textbf {\bibinfo {volume} {22}},\ \bibinfo {pages} {218} (\bibinfo
  {year} {1921})}\BibitemShut {NoStop}%
\bibitem [{\citenamefont {LANDSBERG}(1964)}]{LANDSBERG1964}%
  \BibitemOpen
  \bibfield  {author} {\bibinfo {author} {\bibfnamefont {P.~T.}\ \bibnamefont
  {LANDSBERG}},\ }\href {\doibase 10.1038/201485b0} {\bibfield  {journal}
  {\bibinfo  {journal} {Nature}\ }\textbf {\bibinfo {volume} {201}},\ \bibinfo
  {pages} {485} (\bibinfo {year} {1964})}\BibitemShut {NoStop}%
\bibitem [{\citenamefont {Pogliani}\ and\ \citenamefont
  {Berberan-Santos}(2000)}]{Pogliani2000}%
  \BibitemOpen
  \bibfield  {author} {\bibinfo {author} {\bibfnamefont {L.}~\bibnamefont
  {Pogliani}}\ and\ \bibinfo {author} {\bibfnamefont {M.~N.}\ \bibnamefont
  {Berberan-Santos}},\ }\href {\doibase 10.1023/a:1018834326958} {\bibfield
  {journal} {\bibinfo  {journal} {J. Math. Chem.}\ }\textbf {\bibinfo {volume}
  {28}},\ \bibinfo {pages} {313} (\bibinfo {year} {2000})}\BibitemShut
  {NoStop}%
\bibitem [{\citenamefont {Zoric}(2011)}]{Zoric2011}%
  \BibitemOpen
  \bibfield  {author} {\bibinfo {author} {\bibfnamefont {V.~A.}\ \bibnamefont
  {Zoric}},\ }\href@noop {} {\emph {\bibinfo {title} {Mathematical Analysis of
  Problems in the Natural Sciences}}},\ SpringerLink\ (\bibinfo  {publisher}
  {Springer-Verlag Berlin Heidelberg},\ \bibinfo {address} {Berlin,
  Heidelberg},\ \bibinfo {year} {2011})\ \bibinfo {note} {original Russian ed.:
  Matematicheskij analiz zadach estestvoznaniya published by MCCME, Moscow,
  Russia, 2008}\BibitemShut {NoStop}%
\bibitem [{\citenamefont {Ma}\ \emph {et~al.}(2021)\citenamefont {Ma},
  \citenamefont {Dong}, \citenamefont {Quan},\ and\ \citenamefont
  {Sun}}]{Ma2021}%
  \BibitemOpen
  \bibfield  {author} {\bibinfo {author} {\bibfnamefont {Y.-H.}\ \bibnamefont
  {Ma}}, \bibinfo {author} {\bibfnamefont {H.}~\bibnamefont {Dong}}, \bibinfo
  {author} {\bibfnamefont {H.-T.}\ \bibnamefont {Quan}}, \ and\ \bibinfo
  {author} {\bibfnamefont {C.-P.}\ \bibnamefont {Sun}},\ }\href {\doibase
  10.1016/j.fmre.2020.11.003} {\bibfield  {journal} {\bibinfo  {journal} {FR}\
  }\textbf {\bibinfo {volume} {1}},\ \bibinfo {pages} {6} (\bibinfo {year}
  {2021})}\BibitemShut {NoStop}%
\bibitem [{\citenamefont {Quan}\ \emph {et~al.}(2005)\citenamefont {Quan},
  \citenamefont {Zhang},\ and\ \citenamefont {Sun}}]{Quan2005}%
  \BibitemOpen
  \bibfield  {author} {\bibinfo {author} {\bibfnamefont {H.~T.}\ \bibnamefont
  {Quan}}, \bibinfo {author} {\bibfnamefont {P.}~\bibnamefont {Zhang}}, \ and\
  \bibinfo {author} {\bibfnamefont {C.~P.}\ \bibnamefont {Sun}},\ }\href
  {\doibase 10.1103/physreve.72.056110} {\bibfield  {journal} {\bibinfo
  {journal} {Phys. Rev. E}\ }\textbf {\bibinfo {volume} {72}},\ \bibinfo
  {pages} {056110} (\bibinfo {year} {2005})}\BibitemShut {NoStop}%
\bibitem [{\citenamefont {Quan}\ \emph {et~al.}(2007)\citenamefont {Quan},
  \citenamefont {Liu}, \citenamefont {Sun},\ and\ \citenamefont
  {Nori}}]{Quan2007}%
  \BibitemOpen
  \bibfield  {author} {\bibinfo {author} {\bibfnamefont {H.~T.}\ \bibnamefont
  {Quan}}, \bibinfo {author} {\bibfnamefont {Y.~X.}\ \bibnamefont {Liu}},
  \bibinfo {author} {\bibfnamefont {C.~P.}\ \bibnamefont {Sun}}, \ and\
  \bibinfo {author} {\bibfnamefont {F.}~\bibnamefont {Nori}},\ }\href {\doibase
  10.1103/physreve.76.031105} {\bibfield  {journal} {\bibinfo  {journal} {Phys.
  Rev. E}\ }\textbf {\bibinfo {volume} {76}},\ \bibinfo {pages} {031105}
  (\bibinfo {year} {2007})}\BibitemShut {NoStop}%
\bibitem [{\citenamefont {Hawking}(1975)}]{Hawking1975}%
  \BibitemOpen
  \bibfield  {author} {\bibinfo {author} {\bibfnamefont {S.~W.}\ \bibnamefont
  {Hawking}},\ }\href {\doibase 10.1007/bf02345020} {\bibfield  {journal}
  {\bibinfo  {journal} {Commun. Math. Phys.}\ }\textbf {\bibinfo {volume}
  {43}},\ \bibinfo {pages} {199} (\bibinfo {year} {1975})}\BibitemShut
  {NoStop}%
\bibitem [{\citenamefont {Hawking}(1976)}]{Hawking1976}%
  \BibitemOpen
  \bibfield  {author} {\bibinfo {author} {\bibfnamefont {S.~W.}\ \bibnamefont
  {Hawking}},\ }\href {\doibase 10.1103/physrevd.14.2460} {\bibfield  {journal}
  {\bibinfo  {journal} {Phys. Rev. D}\ }\textbf {\bibinfo {volume} {14}},\
  \bibinfo {pages} {2460} (\bibinfo {year} {1976})}\BibitemShut {NoStop}%
\bibitem [{\citenamefont {Parikh}\ and\ \citenamefont
  {Wilczek}(2000)}]{Parikh2000}%
  \BibitemOpen
  \bibfield  {author} {\bibinfo {author} {\bibfnamefont {M.~K.}\ \bibnamefont
  {Parikh}}\ and\ \bibinfo {author} {\bibfnamefont {F.}~\bibnamefont
  {Wilczek}},\ }\href {\doibase 10.1103/physrevlett.85.5042} {\bibfield
  {journal} {\bibinfo  {journal} {Phys. Rev. Lett.}\ }\textbf {\bibinfo
  {volume} {85}},\ \bibinfo {pages} {5042} (\bibinfo {year}
  {2000})}\BibitemShut {NoStop}%
\bibitem [{\citenamefont {Zhang}\ \emph {et~al.}(2009)\citenamefont {Zhang},
  \citenamefont {Cai}, \citenamefont {You},\ and\ \citenamefont
  {Zhan}}]{Zhang2009}%
  \BibitemOpen
  \bibfield  {author} {\bibinfo {author} {\bibfnamefont {B.}~\bibnamefont
  {Zhang}}, \bibinfo {author} {\bibfnamefont {Q.-y.}\ \bibnamefont {Cai}},
  \bibinfo {author} {\bibfnamefont {L.}~\bibnamefont {You}}, \ and\ \bibinfo
  {author} {\bibfnamefont {M.-s.}\ \bibnamefont {Zhan}},\ }\href {\doibase
  10.1016/j.physletb.2009.03.082} {\bibfield  {journal} {\bibinfo  {journal}
  {Phys. Lett. B}\ }\textbf {\bibinfo {volume} {675}},\ \bibinfo {pages} {98}
  (\bibinfo {year} {2009})}\BibitemShut {NoStop}%
\bibitem [{\citenamefont {Cai}\ \emph {et~al.}(2016)\citenamefont {Cai},
  \citenamefont {Sun},\ and\ \citenamefont {You}}]{Cai2016}%
  \BibitemOpen
  \bibfield  {author} {\bibinfo {author} {\bibfnamefont {Q.-Y.}\ \bibnamefont
  {Cai}}, \bibinfo {author} {\bibfnamefont {C.~P.}\ \bibnamefont {Sun}}, \ and\
  \bibinfo {author} {\bibfnamefont {L.}~\bibnamefont {You}},\ }\href {\doibase
  10.1016/j.nuclphysb.2016.02.030} {\bibfield  {journal} {\bibinfo  {journal}
  {Nucl. Phys. B}\ }\textbf {\bibinfo {volume} {905}},\ \bibinfo {pages} {327}
  (\bibinfo {year} {2016})}\BibitemShut {NoStop}%
\bibitem [{\citenamefont {Dong}\ \emph {et~al.}(2014)\citenamefont {Dong},
  \citenamefont {Cai}, \citenamefont {Liu},\ and\ \citenamefont
  {Sun}}]{Dong2014}%
  \BibitemOpen
  \bibfield  {author} {\bibinfo {author} {\bibfnamefont {H.}~\bibnamefont
  {Dong}}, \bibinfo {author} {\bibfnamefont {Q.-Y.}\ \bibnamefont {Cai}},
  \bibinfo {author} {\bibfnamefont {X.-F.}\ \bibnamefont {Liu}}, \ and\
  \bibinfo {author} {\bibfnamefont {C.-P.}\ \bibnamefont {Sun}},\ }\href
  {\doibase 10.1088/0253-6102/61/3/03} {\bibfield  {journal} {\bibinfo
  {journal} {Commun. Theor. Phys.}\ }\textbf {\bibinfo {volume} {61}},\
  \bibinfo {pages} {289} (\bibinfo {year} {2014})}\BibitemShut {NoStop}%
\bibitem [{\citenamefont {Hellmann}(1933)}]{Hellmann1933}%
  \BibitemOpen
  \bibfield  {author} {\bibinfo {author} {\bibfnamefont {H.}~\bibnamefont
  {Hellmann}},\ }\href {\doibase 10.1007/bf01342053} {\bibfield  {journal}
  {\bibinfo  {journal} {Zeitschrift f\"ur Physik}\ }\textbf {\bibinfo {volume}
  {85}},\ \bibinfo {pages} {180} (\bibinfo {year} {1933})}\BibitemShut
  {NoStop}%
\bibitem [{\citenamefont {Feynman}(1939)}]{Feynman1939}%
  \BibitemOpen
  \bibfield  {author} {\bibinfo {author} {\bibfnamefont {R.~P.}\ \bibnamefont
  {Feynman}},\ }\href {\doibase 10.1103/physrev.56.340} {\bibfield  {journal}
  {\bibinfo  {journal} {Phys. Rev.}\ }\textbf {\bibinfo {volume} {56}},\
  \bibinfo {pages} {340} (\bibinfo {year} {1939})}\BibitemShut {NoStop}%
\bibitem [{SM_()}]{SM_This}%
  \BibitemOpen
  \href@noop {} {}\bibinfo {note} {See Supplementary Materials for detailed
  discussions on the derivation of the EIE, the principle of detailed balance
  versus the structure of probability, the expansion with small temperature
  deviation, and the non-canonical emission for small black holes.}\BibitemShut
  {Stop}%
\bibitem [{\citenamefont {Breuer}(2007)}]{Breuer2007}%
  \BibitemOpen
  \bibfield  {author} {\bibinfo {author} {\bibfnamefont {H.-P.}\ \bibnamefont
  {Breuer}},\ }\href@noop {} {\emph {\bibinfo {title} {The theory of open
  quantum systems}}},\ edited by\ \bibinfo {editor} {\bibfnamefont
  {F.}~\bibnamefont {Petruccione}}\ (\bibinfo  {publisher} {Clarendon},\
  \bibinfo {address} {Oxford},\ \bibinfo {year} {2007})\BibitemShut {NoStop}%
\bibitem [{\citenamefont {Xu}\ \emph {et~al.}(2014)\citenamefont {Xu},
  \citenamefont {Li}, \citenamefont {Liu},\ and\ \citenamefont {Sun}}]{Xu2014}%
  \BibitemOpen
  \bibfield  {author} {\bibinfo {author} {\bibfnamefont {D.~Z.}\ \bibnamefont
  {Xu}}, \bibinfo {author} {\bibfnamefont {S.-W.}\ \bibnamefont {Li}}, \bibinfo
  {author} {\bibfnamefont {X.~F.}\ \bibnamefont {Liu}}, \ and\ \bibinfo
  {author} {\bibfnamefont {C.~P.}\ \bibnamefont {Sun}},\ }\href {\doibase
  10.1103/physreve.90.062125} {\bibfield  {journal} {\bibinfo  {journal} {Phys.
  Rev. E}\ }\textbf {\bibinfo {volume} {90}},\ \bibinfo {pages} {062125}
  (\bibinfo {year} {2014})}\BibitemShut {NoStop}%
\bibitem [{\citenamefont {Richens}\ \emph {et~al.}(2018)\citenamefont
  {Richens}, \citenamefont {Alhambra},\ and\ \citenamefont
  {Masanes}}]{Richens2018}%
  \BibitemOpen
  \bibfield  {author} {\bibinfo {author} {\bibfnamefont {J.~G.}\ \bibnamefont
  {Richens}}, \bibinfo {author} {\bibfnamefont {A.~M.}\ \bibnamefont
  {Alhambra}}, \ and\ \bibinfo {author} {\bibfnamefont {L.}~\bibnamefont
  {Masanes}},\ }\href {\doibase 10.1103/physreve.97.062132} {\bibfield
  {journal} {\bibinfo  {journal} {Phys. Rev. E}\ }\textbf {\bibinfo {volume}
  {97}},\ \bibinfo {pages} {062132} (\bibinfo {year} {2018})}\BibitemShut
  {NoStop}%
\bibitem [{\citenamefont {Chung}\ and\ \citenamefont
  {Hassanabadi}(2019)}]{Chung2019}%
  \BibitemOpen
  \bibfield  {author} {\bibinfo {author} {\bibfnamefont {W.~S.}\ \bibnamefont
  {Chung}}\ and\ \bibinfo {author} {\bibfnamefont {H.}~\bibnamefont
  {Hassanabadi}},\ }\href {\doibase 10.1016/j.physa.2019.121720} {\bibfield
  {journal} {\bibinfo  {journal} {Physica A}\ }\textbf {\bibinfo {volume}
  {532}},\ \bibinfo {pages} {121720} (\bibinfo {year} {2019})}\BibitemShut
  {NoStop}%
\bibitem [{\citenamefont {Beck}\ and\ \citenamefont {Cohen}(2003)}]{Beck2003}%
  \BibitemOpen
  \bibfield  {author} {\bibinfo {author} {\bibfnamefont {C.}~\bibnamefont
  {Beck}}\ and\ \bibinfo {author} {\bibfnamefont {E.}~\bibnamefont {Cohen}},\
  }\href {\doibase 10.1016/s0378-4371(03)00019-0} {\bibfield  {journal}
  {\bibinfo  {journal} {Physica A}\ }\textbf {\bibinfo {volume} {322}},\
  \bibinfo {pages} {267} (\bibinfo {year} {2003})}\BibitemShut {NoStop}%
\bibitem [{\citenamefont {Falcioni}\ \emph {et~al.}(2011)\citenamefont
  {Falcioni}, \citenamefont {Villamaina}, \citenamefont {Vulpiani},
  \citenamefont {Puglisi},\ and\ \citenamefont {Sarracino}}]{Falcioni2011}%
  \BibitemOpen
  \bibfield  {author} {\bibinfo {author} {\bibfnamefont {M.}~\bibnamefont
  {Falcioni}}, \bibinfo {author} {\bibfnamefont {D.}~\bibnamefont
  {Villamaina}}, \bibinfo {author} {\bibfnamefont {A.}~\bibnamefont
  {Vulpiani}}, \bibinfo {author} {\bibfnamefont {A.}~\bibnamefont {Puglisi}}, \
  and\ \bibinfo {author} {\bibfnamefont {A.}~\bibnamefont {Sarracino}},\ }\href
  {\doibase 10.1119/1.3563046} {\bibfield  {journal} {\bibinfo  {journal} {Am.
  J. Phys.}\ }\textbf {\bibinfo {volume} {79}},\ \bibinfo {pages} {777}
  (\bibinfo {year} {2011})}\BibitemShut {NoStop}%
\bibitem [{\citenamefont {Fei}\ and\ \citenamefont {Ma}(2024)}]{Fei2024}%
  \BibitemOpen
  \bibfield  {author} {\bibinfo {author} {\bibfnamefont {Z.}~\bibnamefont
  {Fei}}\ and\ \bibinfo {author} {\bibfnamefont {Y.-H.}\ \bibnamefont {Ma}},\
  }\href {\doibase 10.1103/physreve.109.044101} {\bibfield  {journal} {\bibinfo
   {journal} {Phys. Rev. E}\ }\textbf {\bibinfo {volume} {109}},\ \bibinfo
  {pages} {044101} (\bibinfo {year} {2024})}\BibitemShut {NoStop}%
\bibitem [{\citenamefont {Onsager}(1944)}]{Onsager1944}%
  \BibitemOpen
  \bibfield  {author} {\bibinfo {author} {\bibfnamefont {L.}~\bibnamefont
  {Onsager}},\ }\href {\doibase 10.1103/physrev.65.117} {\bibfield  {journal}
  {\bibinfo  {journal} {Phys. Rev.}\ }\textbf {\bibinfo {volume} {65}},\
  \bibinfo {pages} {117} (\bibinfo {year} {1944})}\BibitemShut {NoStop}%
\bibitem [{\citenamefont {Yang}(1952)}]{Yang1952}%
  \BibitemOpen
  \bibfield  {author} {\bibinfo {author} {\bibfnamefont {C.~N.}\ \bibnamefont
  {Yang}},\ }\href {\doibase 10.1103/physrev.85.808} {\bibfield  {journal}
  {\bibinfo  {journal} {Physical Review}\ }\textbf {\bibinfo {volume} {85}},\
  \bibinfo {pages} {808} (\bibinfo {year} {1952})}\BibitemShut {NoStop}%
\bibitem [{\citenamefont {Fisher}\ and\ \citenamefont
  {Ferdinand}(1967)}]{Fisher1967}%
  \BibitemOpen
  \bibfield  {author} {\bibinfo {author} {\bibfnamefont {M.~E.}\ \bibnamefont
  {Fisher}}\ and\ \bibinfo {author} {\bibfnamefont {A.~E.}\ \bibnamefont
  {Ferdinand}},\ }\href {\doibase 10.1103/physrevlett.19.169} {\bibfield
  {journal} {\bibinfo  {journal} {Phys. Rev. Lett.}\ }\textbf {\bibinfo
  {volume} {19}},\ \bibinfo {pages} {169} (\bibinfo {year} {1967})}\BibitemShut
  {NoStop}%
\bibitem [{\citenamefont {Ferdinand}\ and\ \citenamefont
  {Fisher}(1969)}]{Ferdinand1969}%
  \BibitemOpen
  \bibfield  {author} {\bibinfo {author} {\bibfnamefont {A.~E.}\ \bibnamefont
  {Ferdinand}}\ and\ \bibinfo {author} {\bibfnamefont {M.~E.}\ \bibnamefont
  {Fisher}},\ }\href {\doibase 10.1103/physrev.185.832} {\bibfield  {journal}
  {\bibinfo  {journal} {Phys. Rev.}\ }\textbf {\bibinfo {volume} {185}},\
  \bibinfo {pages} {832} (\bibinfo {year} {1969})}\BibitemShut {NoStop}%
\bibitem [{\citenamefont {Binder}(1972)}]{Binder1972}%
  \BibitemOpen
  \bibfield  {author} {\bibinfo {author} {\bibfnamefont {K.}~\bibnamefont
  {Binder}},\ }\href {\doibase 10.1016/0031-8914(72)90237-6} {\bibfield
  {journal} {\bibinfo  {journal} {Physica}\ }\textbf {\bibinfo {volume} {62}},\
  \bibinfo {pages} {508} (\bibinfo {year} {1972})}\BibitemShut {NoStop}%
\bibitem [{\citenamefont {Privman}\ and\ \citenamefont
  {Fisher}(1984)}]{Privman1984}%
  \BibitemOpen
  \bibfield  {author} {\bibinfo {author} {\bibfnamefont {V.}~\bibnamefont
  {Privman}}\ and\ \bibinfo {author} {\bibfnamefont {M.~E.}\ \bibnamefont
  {Fisher}},\ }\href {\doibase 10.1103/physrevb.30.322} {\bibfield  {journal}
  {\bibinfo  {journal} {Phys. Rev. B}\ }\textbf {\bibinfo {volume} {30}},\
  \bibinfo {pages} {322} (\bibinfo {year} {1984})}\BibitemShut {NoStop}%
\bibitem [{\citenamefont {Wald}(2001)}]{Wald2001}%
  \BibitemOpen
  \bibfield  {author} {\bibinfo {author} {\bibfnamefont {R.~M.}\ \bibnamefont
  {Wald}},\ }\href {\doibase 10.12942/lrr-2001-6} {\bibfield  {journal}
  {\bibinfo  {journal} {Living Rev. Relativ.}\ }\textbf {\bibinfo {volume} {4}}
  (\bibinfo {year} {2001}),\ 10.12942/lrr-2001-6}\BibitemShut {NoStop}%
\bibitem [{\citenamefont {Carlip}(2014)}]{Carlip2014}%
  \BibitemOpen
  \bibfield  {author} {\bibinfo {author} {\bibfnamefont {S.}~\bibnamefont
  {Carlip}},\ }\href {\doibase 10.1142/s0218271814300237} {\bibfield  {journal}
  {\bibinfo  {journal} {Int. J. Mod. Phys. D}\ }\textbf {\bibinfo {volume}
  {23}},\ \bibinfo {pages} {1430023} (\bibinfo {year} {2014})}\BibitemShut
  {NoStop}%
\end{thebibliography}%


\begin{thebibliography}{6}%
\makeatletter
\providecommand \@ifxundefined [1]{%
 \@ifx{#1\undefined}
}%
\providecommand \@ifnum [1]{%
 \ifnum #1\expandafter \@firstoftwo
 \else \expandafter \@secondoftwo
 \fi
}%
\providecommand \@ifx [1]{%
 \ifx #1\expandafter \@firstoftwo
 \else \expandafter \@secondoftwo
 \fi
}%
\providecommand \natexlab [1]{#1}%
\providecommand \enquote  [1]{``#1''}%
\providecommand \bibnamefont  [1]{#1}%
\providecommand \bibfnamefont [1]{#1}%
\providecommand \citenamefont [1]{#1}%
\providecommand \href@noop [0]{\@secondoftwo}%
\providecommand \href [0]{\begingroup \@sanitize@url \@href}%
\providecommand \@href[1]{\@@startlink{#1}\@@href}%
\providecommand \@@href[1]{\endgroup#1\@@endlink}%
\providecommand \@sanitize@url [0]{\catcode `\\12\catcode `\$12\catcode `\&12\catcode `\#12\catcode `\^12\catcode `\_12\catcode `\%12\relax}%
\providecommand \@@startlink[1]{}%
\providecommand \@@endlink[0]{}%
\providecommand \url  [0]{\begingroup\@sanitize@url \@url }%
\providecommand \@url [1]{\endgroup\@href {#1}{\urlprefix }}%
\providecommand \urlprefix  [0]{URL }%
\providecommand \Eprint [0]{\href }%
\providecommand \doibase [0]{http://dx.doi.org/}%
\providecommand \selectlanguage [0]{\@gobble}%
\providecommand \bibinfo  [0]{\@secondoftwo}%
\providecommand \bibfield  [0]{\@secondoftwo}%
\providecommand \translation [1]{[#1]}%
\providecommand \BibitemOpen [0]{}%
\providecommand \bibitemStop [0]{}%
\providecommand \bibitemNoStop [0]{.\EOS\space}%
\providecommand \EOS [0]{\spacefactor3000\relax}%
\providecommand \BibitemShut  [1]{\csname bibitem#1\endcsname}%
\let\auto@bib@innerbib\@empty
\bibitem [{\citenamefont {Breuer}(2007)}]{Breuer2007}%
  \BibitemOpen
  \bibfield  {author} {\bibinfo {author} {\bibfnamefont {Heinz-Peter}\ \bibnamefont {Breuer}},\ }\href@noop {} {\emph {\bibinfo {title} {The theory of open quantum systems}}},\ edited by\ \bibinfo {editor} {\bibfnamefont {F.}~\bibnamefont {Petruccione}}\ (\bibinfo  {publisher} {Clarendon},\ \bibinfo {address} {Oxford},\ \bibinfo {year} {2007})\BibitemShut {NoStop}%
\bibitem [{\citenamefont {Carath\'eodory}(1909)}]{Caratheodory1909}%
  \BibitemOpen
  \bibfield  {author} {\bibinfo {author} {\bibfnamefont {C.}~\bibnamefont {Carath\'eodory}},\ }\bibfield  {title} {\enquote {\bibinfo {title} {{U}ntersuchungen \"uber die {G}rundlagen der {T}hermodynamik},}\ }\href {\doibase 10.1007/bf01450409} {\bibfield  {journal} {\bibinfo  {journal} {Math. Ann.}\ }\textbf {\bibinfo {volume} {67}},\ \bibinfo {pages} {355--386} (\bibinfo {year} {1909})}\BibitemShut {NoStop}%
\bibitem [{\citenamefont {Born}(1921)}]{Born1921}%
  \BibitemOpen
  \bibfield  {author} {\bibinfo {author} {\bibfnamefont {Max}\ \bibnamefont {Born}},\ }\bibfield  {title} {\enquote {\bibinfo {title} {{K}ritische betrachtungen zur traditionellen darstellung der thermodynamik},}\ }\href@noop {} {\bibfield  {journal} {\bibinfo  {journal} {Phys. Zeit.}\ }\textbf {\bibinfo {volume} {22}},\ \bibinfo {pages} {218--224} (\bibinfo {year} {1921})}\BibitemShut {NoStop}%
\bibitem [{\citenamefont {Yang}(1952)}]{Yang1952}%
  \BibitemOpen
  \bibfield  {author} {\bibinfo {author} {\bibfnamefont {C.~N.}\ \bibnamefont {Yang}},\ }\bibfield  {title} {\enquote {\bibinfo {title} {The spontaneous magnetization of a two-dimensional ising model},}\ }\href {\doibase 10.1103/physrev.85.808} {\bibfield  {journal} {\bibinfo  {journal} {Physical Review}\ }\textbf {\bibinfo {volume} {85}},\ \bibinfo {pages} {808--816} (\bibinfo {year} {1952})}\BibitemShut {NoStop}%
\bibitem [{\citenamefont {Privman}\ and\ \citenamefont {Fisher}(1984)}]{Privman1984}%
  \BibitemOpen
  \bibfield  {author} {\bibinfo {author} {\bibfnamefont {Valdimir}\ \bibnamefont {Privman}}\ and\ \bibinfo {author} {\bibfnamefont {Michael~E.}\ \bibnamefont {Fisher}},\ }\bibfield  {title} {\enquote {\bibinfo {title} {Universal critical amplitudes in finite-size scaling},}\ }\href {\doibase 10.1103/physrevb.30.322} {\bibfield  {journal} {\bibinfo  {journal} {Phys. Rev. B}\ }\textbf {\bibinfo {volume} {30}},\ \bibinfo {pages} {322--327} (\bibinfo {year} {1984})}\BibitemShut {NoStop}%
\bibitem [{\citenamefont {Parikh}\ and\ \citenamefont {Wilczek}(2000)}]{Parikh2000}%
  \BibitemOpen
  \bibfield  {author} {\bibinfo {author} {\bibfnamefont {Maulik~K.}\ \bibnamefont {Parikh}}\ and\ \bibinfo {author} {\bibfnamefont {Frank}\ \bibnamefont {Wilczek}},\ }\bibfield  {title} {\enquote {\bibinfo {title} {{H}awking {R}adiation as {T}unneling},}\ }\href {\doibase 10.1103/physrevlett.85.5042} {\bibfield  {journal} {\bibinfo  {journal} {Phys. Rev. Lett.}\ }\textbf {\bibinfo {volume} {85}},\ \bibinfo {pages} {5042--5045} (\bibinfo {year} {2000})}\BibitemShut {NoStop}%
\end{thebibliography}%

\end{document}


\title{Supplementary Materials of ``Quantum Thermodynamic Integrability
for Canonical and non-Canonical Statistics''}
\author{R.X. Zhai}
\affiliation{Graduate School of China Academy of Engineering Physics,~\\
No.10 Xibeiwang East Road, Haidian District, Beijing 100193, China}
\author{C.P. Sun}
\email{suncp@gscaep.ac.cn}

\affiliation{Graduate School of China Academy of Engineering Physics,~\\
No.10 Xibeiwang East Road, Haidian District, Beijing 100193, China}
\date{\today}

\maketitle
These supplementary materials provide detailed derivations and discussions
related to the main text. In Sec. \ref{sec:Derivation-EIE}, the entropy
integrable equations in different forms are derived. In Sec. \ref{sec:detial_balance},
we begin with the principle of detailed balance and determine the
Boltzmann distribution using Quantum Thermodynamic Integrability (QTI).
In Sec. \ref{sec:Construction-of-non-canonical} we construct a non-canonical
solution to the entropy integrable equations (EIE) by linearly combining
the canonical states at varying temperatures. In Sec. \ref{sec:Expansion-with-small},
the quantities of non-canonical solutions of EIE are expanded for
a small deviation in the inverse temperature. In Sec. \ref{sec:non_canonical_Ising},
the thermodynamic functions of non-canonical states of the two-dimensional
Ising model are evaluated. In Sec. \ref{sec:non_can_correlation},
the information correlation for non-canonical states is discussed.
In Sec. \ref{sec:Emission-Rate-for}, the non-canonical radiation
from a small black hole is discussed with the non-canonical solution
to the EIE.

\section{Derivation of the entropy integrable equations}

\label{sec:Derivation-EIE}

\subsection{QTI from the mechanical parameters to the energy levels}

We consider a quantum system whose Hamiltonian $H(\bm{\lambda})$
is a function of $K$ independent macroscopic mechanical parameters
$\bm{\lambda}\equiv(\lambda_{1},\lambda_{2},\cdots,\lambda_{K})$.
The energy levels $E_{n}(\bm{\lambda})$ and the eigenstates $\left|n(\bm{\lambda})\right\rangle $
are determined as $H(\bm{\lambda})\left|n\right\rangle =E_{n}\left|n\right\rangle $.
The density matrix of this system is determined by $\bm{\lambda}$
and a thermodynamic parameter $\xi$ to be determined, i.e., $\rho=\rho(\bm{\lambda},\xi)$.
On the other hand, the heat for such a system within the parameterized
TDM: $\{(\bm{\lambda},\xi)\}$ is expressed as
\[
\delta Q=\mathrm{tr}\left(H\mathrm{d}\rho\right)=\mathrm{tr}\left(H\partial_{\mu}\rho\right)\mathrm{d}\lambda_{\mu}+\mathrm{tr}\left(H\partial_{\xi}\rho\right)\mathrm{d}\xi.
\]
The condition of QTI gives
\begin{align}
\mathrm{d}\left(\beta\delta Q\right)=\beta\mathrm{d}\left(\delta Q\right)+\mathrm{d}\beta\wedge\delta Q & =\beta\mathrm{tr}\left(\partial_{\nu}H\partial_{\mu}\rho\right)\mathrm{d}\lambda_{\nu}\wedge\mathrm{d}\lambda_{\mu}+\beta\mathrm{tr}\left(\partial_{\nu}H\partial_{\xi}\rho\right)\mathrm{d}\lambda_{\nu}\wedge\mathrm{d}\xi\label{eq:QTI_Condi_xi}\\
+\mathrm{tr}\left(H\partial_{\mu}\rho\right)\mathrm{d}\beta\wedge\mathrm{d}\lambda_{\mu}+ & \mathrm{tr}\left(H\partial_{\xi}\rho\right)\mathrm{d}\beta\wedge\mathrm{d}\xi=0.\nonumber 
\end{align}
We take the simplest integrating factor as $\beta=\xi$, which is
independent of the mechanical parameters $\bm{\lambda}$. In this
situation, the condition of QTI in Eq. \eqref{eq:QTI_Condi_xi} simplifies
into
\begin{equation}
\sum_{\mu<\nu}\mathrm{tr}\left(\partial_{\nu}H\partial_{\mu}\rho-\partial_{\mu}H\partial_{\nu}\rho\right)\mathrm{d}\lambda_{\nu}\wedge\mathrm{d}\lambda_{\mu}+\sum_{\mu}\mathrm{tr}\left(\beta\partial_{\mu}H\partial_{\beta}\rho-H\partial_{\mu}\rho\right)\mathrm{d}\lambda_{\mu}\wedge\mathrm{d}\beta=0.\label{eq:QTI_condition_p_TDM}
\end{equation}
This gives Eqs. (2)-(3) in the main text.

For equilibrium states, we have $\left[\rho,H\right]=0$, i.e., the
density matrix should be diagonalized as $\rho=\sum_{n}P_{n}(\bm{\lambda},\beta)\left|n\right\rangle \left\langle n\right|$.
As a result, it follows that
\[
\mathrm{tr}\left(\partial_{\nu}H\partial_{\mu}\rho\right)=\sum_{m,n}\left\langle m\right|\partial_{\nu}H\left|n\right\rangle \left\langle n\right|\partial_{\mu}\rho\left|m\right\rangle =\sum_{n}\left(\partial_{\mu}E_{n}\right)\left(\partial_{\nu}P_{n}\right)+\sum_{m,n}\left(E_{n}-E_{m}\right)\left(P_{n}-P_{m}\right)\langle m\vert\partial_{\mu}n\rangle\langle\partial_{\nu}n|m\rangle.
\]
We can rewrite the condition of QTI in Eq. \eqref{eq:QTI_condition_p_TDM}
into: 
\begin{equation}
\mathrm{tr}\left(\partial_{\nu}H\partial_{\mu}\rho-\partial_{\mu}H\partial_{\nu}\rho\right)=\sum_{n}\left(\partial_{\nu}E_{n}\partial_{\mu}P_{n}-\partial_{\mu}E_{n}\partial_{\nu}P_{n}\right)=0,\label{eq:macro_diagonlized_QTI_01}
\end{equation}
and 
\begin{equation}
\mathrm{tr}\left(\beta\partial_{\mu}H\partial_{\beta}\rho-H\partial_{\mu}\rho\right)=\sum_{n}\left(E_{n}\partial_{\mu}P_{n}-\beta\partial_{\mu}E_{n}\partial_{\beta}P_{n}\right)=0.\label{eq:macro_diagonlized_QTI_02}
\end{equation}
By assuming that $P_{n}$ relies on $\bm{\lambda}$ through $\bm{E}(\bm{\lambda})$,
we have $\partial_{\mu}=\sum_{m}\left(\partial_{\mu}E_{m}\right)\partial_{m}$,
and Eqs. \eqref{eq:macro_diagonlized_QTI_01} and \eqref{eq:macro_diagonlized_QTI_02}
become Eqs. (4) and (5) in the main text.

On the other hand, the QTI condition within the TDM: $\{(\bm{E},\xi)\}$
reads
\[
\mathrm{d}\left(\beta\delta Q\right)=\sum_{m<n}\left(\partial_{m}P_{n}-\partial_{n}P_{m}\right)\mathrm{d}E_{n}\wedge\mathrm{d}E_{m}+\sum_{n}\left(\sum_{m}E_{m}\partial_{n}P_{m}-\beta\partial_{\beta}P_{n}\right)\mathrm{d}\beta\wedge\mathrm{d}E_{n}=0.
\]
These equations are exactly the EIE (Eqs. (6) and (7) in the main
text), and sufficiently guarantee the QTI condition expressed with
Eqs. (4) and (5) in the main text. Namely, we conclude that the QTI
within (macroscopic) parameterized TDM is sufficiently guaranteed
by the QTI within the (microscopic) TDM.

\subsection{Entropy integrable equation in the form of generalized free energy}

Eq. (6) in the main text allows us to define a function $F(\beta,\bm{E})$
with $\bm{P}=\bm{\nabla}F$. Now we express Eq. (7) in the main text
using $F$, resulting in
\[
(\bm{E}\cdot\bm{\nabla})\bm{\nabla}F-\beta\partial_{\beta}\bm{\nabla}F=\bm{\nabla}\left[(\bm{E}\cdot\bm{\nabla})F-\beta\partial_{\beta}F\right]=0.
\]
This means the term inside the bracket is a function of $\beta$,
being independent of $\bm{E}$, namely, $(\bm{E}\cdot\bm{\nabla})F-F-\beta\partial_{\beta}F=A(\beta).$
By applying a transformation $F\rightarrow F+B(\beta)$ to the generalized
free energy, we obtain the EIE in the form of free energy (Eq. (8)
in the main text) as 
\begin{equation}
(\bm{E}\cdot\bm{\nabla})F-F-\beta\partial_{\beta}F=0,
\end{equation}
where $B(\beta)$ is determined by $B+B'=-A$. It can be verified
that the transformation does not change the probability distribution
$\bm{P}$.

\section{Equilibrium distribution from the EIE}

\label{sec:detial_balance} In this section, the principle of detailed
balance, as mentioned in the main text, is discussed in detail.

\subsection{Microscopic mechanism of the principle of detailed balance}

Consider a system with a Hamiltonian $H_{S}=\sum_{n}E_{n}\left|n\right\rangle \left\langle n\right|$,
and the transition term induced by the reservoir $V=\sum_{m,n}V_{mn}\left|m\right\rangle \left\langle n\right|$.
We suppose that the quantum state reads $\left|\psi(t)\right\rangle =\sum_{n}c_{n}(t)\mathrm{e}^{-\mathrm{i}E_{n}t}\left|n\right\rangle $.
Using the Schrödinger Equation, the equation of motion for $c_{n}(t)$
is obtained as 
\[
\mathrm{i}\dot{c}_{n}(t)=\sum_{m}V_{nm}\mathrm{e}^{\mathrm{i}(E_{n}-E_{m})t}\left|n\right\rangle .
\]
We can then write the transition of level $n$ as, 
\[
c_{n}(t)-c_{n}(0)=-\mathrm{i}\sum_{m}T_{nm}(t)c_{m}(0).
\]
Within the coherence time, the transition amplitude $T_{nm}(t)$ is
obtained as a series to the order of $V$, 
\[
T_{nm}(t)=V_{nm}\int_{0}^{t_{0}}\mathrm{d}t_{1}\mathrm{e}^{\mathrm{i}(E_{n}-E_{m})t_{1}}-\mathrm{i}\sum_{k}V_{nk}V_{km}\int_{0}^{t_{0}}\mathrm{d}t_{1}\mathrm{e}^{\mathrm{i}(E_{n}-E_{m})t_{1}}\int_{0}^{t_{1}}\mathrm{d}t_{2}\mathrm{e}^{\mathrm{i}(E_{k}-E_{m})t_{2}}.
\]
To the first order of $V$, the transition amplitude between $m$
and $n$ does not depend on any intermediate levels, which implies
that the transition rate $W_{m\rightarrow n}$ between $m$ and $n$
is not dependent on energies of any other levels $k\neq m,n$. Namely,
we can write the transition between levels as $\dot{P}_{n}=\sum_{m}W_{m\rightarrow n}(E_{m},E_{n})P_{m}$.
This discussion is effective only for the situation where the interaction
term $V$ is weak.

\subsection{Structure of equilibrium distribution determined by the principle
of detailed balance}

According to the principle of detailed balance, the rate of each transition
process between levels $m$ and $n$ exactly balances the rate of
its inverse process \citep{Breuer2007}. When the transition rate
$W_{m\rightarrow n}(E_{m},E_{n})$ from $m$ to $n$ is independent
of the energy of other levels $E_{k}$ for any $k\neq m,n$, the principle
of detailed balance is expressed as
\[
P_{m}W_{m\rightarrow n}(E_{m},E_{n})=P_{n}W_{n\rightarrow m}(E_{n},E_{m}).
\]
Thus, the relative probability $P_{m}/P_{n}$ between levels $m$
and $n$ is also independent of $E_{k}$ for any $k\neq m,n$. We
denote its logarithm by $\ln(P_{m}/P_{n})=\ln P_{m}-\ln P_{n}\equiv f_{mn}(E_{m},E_{n}).$

For arbitrary $m,n\neq1$, by inserting a $\ln P_{1}$ into $f_{mn}(E_{m},E_{n})$,
a `chain rule' is obtained for $f_{mn}$ as $f_{mn}(E_{m},E_{n})=f_{m1}(E_{m},E_{1})-f_{n1}(E_{n},E_{1})$.
Considering the fact that the function $f_{mn}(E_{m},E_{n})$ is independent
of the value of $E_{1}$, the equation holds for $E_{1}=0$,
\begin{equation}
f_{mn}(E_{m},E_{n})=g_{m}(E_{m})-g_{n}(E_{n}),\label{eq:f_mn_fact}
\end{equation}
where $g_{n}(E_{n})\equiv f_{n1}(E_{n},0)$ for $n\neq1$. Furthermore,
for the situation with $n=1$, we have
\[
f_{m1}(E_{m},E_{1})=f_{mk}(E_{m},E_{k})-f_{1k}(E_{1},E_{k})=g_{m}(E_{m})-g_{k}(E_{k})-f_{1k}(E_{1},E_{k}),
\]
which is independent of $k\neq1$ and the value of $E_{k}$. Namely,
we can define that
\[
g_{2}(E_{2})+f_{12}(E_{1},E_{2})=\cdots=g_{k}(E_{k})+f_{1k}(E_{1},E_{k})=\cdots\equiv g_{1}(E_{1}).
\]
Thus, Eq. (\ref{eq:f_mn_fact}) holds for all $m,n$. Denoting $\exp(g_{m}(E_{m}))$
as $\omega_{m}(E_{m})$, the relative probability is re-expressed
as
\begin{equation}
\frac{P_{m}}{P_{n}}=\frac{\omega_{m}(E_{m})}{\omega_{n}(E_{n})}.\label{eq:rel_prob_ratio}
\end{equation}
Summing Eq. (\ref{eq:rel_prob_ratio}) over $m$, we obtain 
\[
P_{n}=\frac{\omega_{n}(E_{n})}{\sum_{m}\omega_{m}(E_{m})}.
\]

\subsection{Boltzmann distribution as unique solution to the EIE}

The ansatz of symmetric generalized free energy in the main text reads
$F=\psi\left[\sum_{n=1}^{N}w(\beta E_{n})\right]/\beta$, for which
the EIE has already been satisfied. To guarantee the normalization
condition $\bm{e}\cdot\bm{\nabla}F=1$, it follows as $\psi'\left[\sum_{n=1}^{N}w(\beta E_{n})\right]=\left(\sum_{n}w'(\beta E_{n})\right)^{-1}$.
Thus, the distribution is in the form
\[
P_{n}=\frac{w'(\beta E_{n})}{\sum_{m}w'(\beta E_{m})}.
\]
Where $w'$ and $\psi'$ are the derivatives of the functions $w$
and $\psi$. The partial derivatives commute, i.e., $\partial_{m}\partial_{n}F=\partial_{n}\partial_{m}F$,
providing an equation related to $w$ as
\[
\frac{w''(\beta E_{n})}{w'(\beta E_{n})}=\frac{w''(\beta E_{m})}{w'(\beta E_{m})}.
\]
The left-hand side is only a function of $\beta E_{n}$, while the
right-hand side is only a function of $\beta E_{m}$. The equation
establishes that both of them equal to a universal constant $-c$
for all $m,n$, regardless of $\beta E_{m}$ and $\beta E_{n}$. Thus,
it reads $w''=-cw'$, whose solution is $w(\beta E_{n})=\exp\left[-c\beta E_{n}\right]$.
Since the entropy never decreases, which can be sufficiently determined
with one experiment, the values of $c$ and $\beta$ are naturally
set to be positive \citep{Caratheodory1909,Born1921}. The discussions
above determine $P_{n}$ to the Boltzmann distribution, up to a linear
rescaling of the inverse temperature $\beta\rightarrow c\beta$. This
physically represents the freedom of choosing the unit of the temperature.
Without loss of generality, we choose $c=1$, inducing the Boltzmann
distribution
\[
P^{(0)}=\frac{1}{Z}\mathrm{e}^{-\beta E_{n}},\quad Z=\sum_{m}\mathrm{e}^{-\beta E_{m}}.
\]
Correspondingly, we can obtain $w(\beta E_{n})=\mathrm{e}^{-\beta E_{n}}$
and $\psi(W)=-\ln(W)$. Therefore, the equilibrium free energy is
given by $F_{0}=-\left(1/\beta\right)\ln Z$.

\section{Construction of non-canonical states}

\label{sec:Construction-of-non-canonical}

A non-canonical state is constructed with $F(\beta,\bm{E})=\int_{0}^{\infty}f(\beta,\beta')F_{0}(\beta,\bm{E}).$
We explore the condition for $f(\beta,\beta')$ to satisfy the EIE,
with $\beta$ being the integrating factor of heat. It follows from
Eq. (8) in the main text that
\begin{equation}
\bm{E}\cdot\bm{\nabla}\int_{0}^{\infty}f(\beta,\beta')F_{0}(\beta',\bm{E})\mathrm{d}\beta=\partial_{\beta}\left(\beta\int_{0}^{\infty}F_{0}(\beta',\bm{E})f(\beta,\beta')\mathrm{d}\beta'\right).\label{eq:EIE_free_energy}
\end{equation}
Integrating the left-hand side (LHS) of Eq. \eqref{eq:EIE_free_energy}
by parts, we obtain
\[
\text{LHS}=\int_{0}^{\infty}f(\beta,\beta')\partial_{\beta'}(\beta'F_{0}(\beta',\bm{E}))\mathrm{d}\beta'=-\int_{0}^{\infty}\beta'F_{0}(\beta',\bm{E})\partial_{\beta'}f(\beta,\beta')\mathrm{d}\beta'.
\]
Here, we have used the conditions $f(\beta,\beta')\rightarrow0$ for
$\beta'\rightarrow0$ and $\beta'\rightarrow\infty$. The right-hand
side (RHS) of Eq. \eqref{eq:EIE_free_energy} reads
\[
\text{RHS}=\int_{0}^{\infty}F_{0}(\beta',\bm{E})\partial_{\beta}\left[\beta f(\beta,\beta')\right]\mathrm{d}\beta'.
\]
Thus, Eq. \eqref{eq:EIE_free_energy} provides
\[
\int_{0}^{\infty}F_{0}(\beta',\bm{E})\left\{ \partial_{\beta}\left[\beta f(\beta,\beta')\right]+\beta'\partial_{\beta'}f(\beta,\beta')\right\} \mathrm{d}\beta'=0,
\]
whose simplest solution reads
\[
\partial_{\beta}\left[\beta f(\beta,\beta')\right]+\beta'\partial_{\beta'}f(\beta,\beta')=0.
\]
This means that $f(\beta',\beta)=g(\beta'/\beta)/\beta$, with $g(x)$
being a normalized probability distribution function. Without loss
of generality, we select the function $g(x)$ as a Gaussian distribution
$\mathcal{N}(1,\sigma^{2})$. Therefore $f(\beta',\beta)$ is obtained
as
\begin{equation}
f(\beta',\beta)=\frac{1}{\Delta\beta\sqrt{2\pi}}\exp\left[-\frac{(\beta'-\beta)^{2}}{2\Delta\beta^{2}}\right],\label{eq:Gaussian_f}
\end{equation}
where $\Delta\beta=\sigma\beta$ is the standard deviation of the
inverse temperature, and is proportional to $\beta$.

\section{Expansion with small temperature deviation}

\label{sec:Expansion-with-small}

We consider an arbitrary quantity $X(\beta')$ as a function of the
inverse temperature. Referring to the non-canonical solution of the
EIE, $\beta'$ is distributed around an average $\beta$ with $f(\beta,\beta')$.
The average of $X$ for the non-canonical state is obtained as 
\[
X^{(\mathrm{nc})}(\beta)=\int_{0}^{\infty}f(\beta,\beta')X(\beta')\mathrm{d}\beta'.
\]
When the standard deviation of $\beta'$ is sufficiently small, we
can expand $X(\beta')$ near the point $\beta'=\beta$, i.e.,
\[
X(\beta')\approx X(\beta)+\left(\beta'-\beta\right)\partial_{\beta}X(\beta)+\frac{1}{2}\left(\beta'-\beta\right)^{2}\partial_{\beta}^{2}X(\beta)+\cdots.
\]
Therefore, $X^{(\mathrm{nc})}(\beta)$ with small deviation can be
expanded to
\begin{equation}
X^{(\mathrm{nc})}(\beta)\approx X(\beta)+\frac{1}{2}\Delta\beta^{2}\partial_{\beta}^{2}X(\beta),\label{eq:general_expansion}
\end{equation}
where we have used$\int_{0}^{\infty}\left(\beta-\beta'\right)f(\beta,\beta')\mathrm{d}\beta'=0'$,
and $\Delta\beta^{2}\equiv\int\left(\beta-\beta'\right)^{2}f(\beta,\beta')\mathrm{d}\beta$.
With this method, the probability of the non-canonical distribution
is obtained as
\begin{equation}
P_{n}^{(\mathrm{nc})}\approx\left(1+a_{n}\Delta\beta^{2}+o(\Delta\beta^{2})\right)P_{n},\label{eq:nc_dist_without_exp}
\end{equation}
where $a_{n}\equiv\left[\left(E_{n}-U\right)^{2}-\left\langle \Delta E^{2}\right\rangle \right]/2$,
with $U$ being the equilibrium internal energy and $\left\langle \Delta E^{k}\right\rangle $
as the $k$-th order central moment of energy. For $a_{n}\Delta\beta^{2}\ll1$,
the non-canonical probability is obtained as 
\[
P_{n}^{(\mathrm{nc})}\approx\frac{\mathrm{e}^{-\beta E_{n}}}{Z}\exp(a_{n}\Delta\beta^{2}),
\]
giving the result in Eq. (11) of the main text. The generalized free
energy, internal energy and thermodynamic entropy can be evaluated
with this method.

As for the relative entropy between $P_{n}$ and $P_{n}^{(\mathcal{\mathrm{nc}})}$,
its leading order is proven to be
\[
D_{KL}(P_{n}||P_{n}^{(\mathrm{nc})})\equiv-\sum_{n}P_{n}\ln\frac{P_{n}^{(\mathrm{nc})}}{P_{n}}\approx\frac{\sigma^{4}}{2}\left\langle a_{n}^{2}\right\rangle =\frac{\Delta\beta^{4}}{8}\left[\left\langle \Delta E^{4}\right\rangle -\left\langle \Delta E^{2}\right\rangle ^{2}\right].
\]
The information entropy can be evaluated with Eq. \eqref{eq:nc_dist_without_exp}.
The quantities with small temperature deviations are listed in Table.
\ref{tab:Quantities-for-non-canonical}.

\begin{table}
\caption{\label{tab:Quantities-for-non-canonical}Quantities for non-canonical
states as functions of $\Delta\beta$ and the equilibrium quantities.}

\begin{spacing}{2.3}
\centering{}\includegraphics{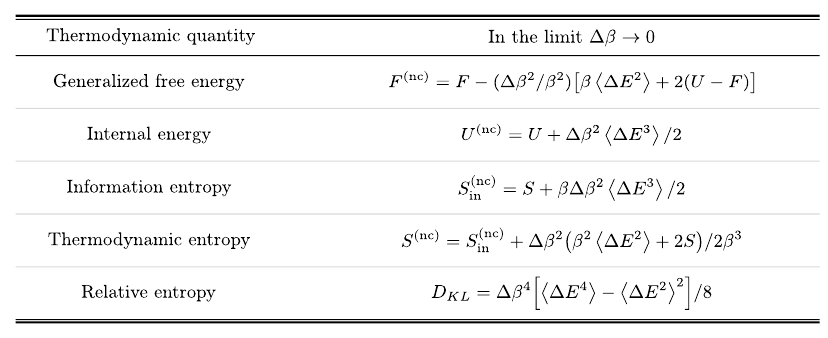}
\end{spacing}
\end{table}

\section{Ising model and non-canonical states}

\label{sec:non_canonical_Ising}

In this section, we discuss the non-canonical properties of the two-dimensional
Ising model.

\subsection{Magnetism and heat capacity for non-canonical distribution}

The Hamiltonian of a two-dimensional Ising model is given by
\[
H(\{s_{i}\})=-J\sum_{\left\langle i,j\right\rangle }s_{i}s_{j},
\]
where the sum over $\left\langle i,j\right\rangle $ represents the
sum over nearest-neighbor lattice sites. The magnetization serves
as the order parameter of the system, defined as 
\[
M(\beta)=\sum_{\{s_{i}\}}P(\{s_{i}\})\frac{\sum_{i}s_{i}}{N}.
\]
Here, $P(\{s_{i}\})$ is the probability distribution of the system.
For canonical states with $P(\beta,\{s_{i}\})=\exp(-\beta H(\{s_{i}\}))/Z$,
there are two different phases of this model. For $\beta>\beta_{c}$
the system shows spontaneous magnetization with $M\neq0$, and for
$\beta<\beta_{c}$, the system shows no magnetism with $M=0$. The
function $M(\beta)$ is analytically obtained as \citep{Yang1952}
\[
M(\beta)=\begin{cases}
0, & \beta<\beta_{c},\\
(1-\sinh^{-4}(2\beta J))^{1/8}, & \beta>\beta_{c}.
\end{cases}
\]
Here, the critical temperature is $\beta_{c}=\ln(1+\sqrt{2})/2J$.

When the system is in non-canonical states with $P^{(\mathrm{nc})}(\beta,\{s_{i}\})=\int\mathrm{d}\beta f(\beta,\beta')P(\beta',\{s_{i}\})$,
the corresponding magnetization is an average of that in canonical
states over different $\beta$, namely
\begin{align}
M^{(\mathrm{nc})}(\beta) & =\sum_{\{s_{i}\}}P^{(\mathrm{nc})}(\beta,\{s_{i}\})\frac{\sum_{i}s_{i}}{N}=\sum_{\{s_{i}\}}\int\mathrm{d}\beta'f(\beta,\beta')P(\beta',\{s_{i}\})\frac{\sum_{i}s_{i}}{N}\nonumber \\
 & =\int\mathrm{d}\beta'f(\beta',\beta)M(\beta').\label{eq:Gaussion_M_nc_sup}
\end{align}
With \eqref{eq:Gaussian_f}, we can numerically obtain the magnetization
for non-canonical Ising model.

The heat capacity $C$ of the two-dimensional Ising model in canonical
states reads
\[
C(\beta)=\frac{4}{\pi}(\beta J\coth2\beta J)^{2}\left\{ \mathrm{K}_{1}(q)-\mathrm{E}_{1}(q)-\left(1-\tanh^{2}(2\beta J)\right)\left[\frac{\pi}{2}+\left(2\tanh^{2}(2\beta J)-1\right)\mathrm{K}_{1}(q)\right]\right\} ,
\]
where $\mathrm{K}_{1}(q)$ is the complete elliptic integral of the
first kind, $\mathrm{E}_{1}(q)$ is the complete elliptic integral
of the second kind, and $q(\beta J)\equiv2\sinh(2\beta J)/\cosh^{2}(2\beta J)$.
When $\beta\rightarrow\beta_{c}$, the heat capacity diverges as $C(\beta)\propto-\ln\left|\beta-\beta_{c}\right|$.
For non-canonical states, the heat capacity reads as follows:
\[
C^{(\mathrm{nc})}(\beta)\equiv-\beta\frac{\partial S^{(\mathrm{nc})}}{\partial\beta}=\int_{0}^{\beta}\mathrm{d}\beta'f(\beta',\beta)C(\beta')\beta/\beta',
\]
 which does not diverge near $\beta=\beta_{c}$. In the numerical
calculation of FIG. 3 in the main text, we used Eq. \eqref{eq:Gaussian_f}
to generate the non-canonical states.

\subsection{Scaling for non-canonical distribution}

With Eqs. \eqref{eq:Gaussian_f} and \eqref{eq:Gaussion_M_nc_sup},
the non-canonical magnetization near $\beta=\beta_{c}$ reads
\begin{align*}
M^{\mathrm{(nc)}} & =\frac{1}{\sqrt{2\pi}}\frac{1}{\sigma\beta}\int_{\beta_{c}}^{\infty}\mathrm{d}\beta'M(\beta')\exp\left\{ -\frac{(\beta'-\beta)^{2}}{2\sigma^{2}\beta^{2}}\right\} \\
 & \propto\frac{(1+t)}{\sigma\beta_{c}}\int_{\beta_{c}}^{\infty}\mathrm{d}\beta'(\beta'-\beta_{c})^{1/8}\exp\left\{ -\frac{[\beta'-\beta_{c}/(1+t)]^{2}}{2\sigma^{2}\beta_{c}^{2}/(1+t)^{2}}\right\} .
\end{align*}
Here $t\equiv(T-T_{c})/T_{c}=(\beta_{c}-\beta)/\beta$, or $\beta=\beta_{c}/(1+t)$.
By assuming $\delta\equiv\beta'-\beta_{c}$, we have
\begin{equation}
M^{\mathrm{(nc)}}\propto\frac{1+t}{\sigma\beta_{c}}\int_{0}^{\infty}\mathrm{d}\delta\delta^{1/8}\exp\left\{ -\frac{[(1+t)\delta+\beta_{c}t]^{2}}{2\sigma^{2}\beta_{c}^{2}}\right\} .\label{eq:M_nc_all_order}
\end{equation}
We only consider the situation with small a temperature deviation,
$\sigma\ll1$, and with the average temperature near the critical
point, $t\ll1$. Since $\sigma$ is small, we take a rather rough
approximation where only the values with $\delta$ near the $\delta_{0}=-\beta_{c}t/(1+t)$
contribute to the integral. Consequently, $\delta$ is also a small
value. The leading order of Eq. \eqref{eq:M_nc_all_order} is then
obtained as
\begin{equation}
M^{\mathrm{(nc)}}\propto\frac{1}{\sigma\beta_{c}}\int_{0}^{\infty}\mathrm{d}\delta\delta^{1/8}\exp\left\{ -\frac{(\delta+\beta_{c}t)^{2}}{2\sigma^{2}\beta_{c}^{2}}\right\} =\ell^{-1/8}f_{M}(\ell t),\label{eq:non_can_scaling}
\end{equation}
where $\ell\equiv1/\sigma$. In Eq. \eqref{eq:non_can_scaling}, we
have omitted the terms including $t\delta$ and $t\delta^{1/8}$,
and
\[
f_{M}(x)\equiv\beta_{c}^{1/8}\int_{0}^{\infty}y^{1/8}\exp\left\{ -(y+x)^{2}/2\right\} \mathrm{d}y.
\]

On the other hand, according to the finite-size scaling theory \citep{Privman1984},
the magnetization of a finite-size two-dimensional Ising model scales
as follows:
\begin{equation}
M^{\mathrm{(finite)}}\propto L^{-1/8}\tilde{f}_{M}(Lt),\label{eq:finite_size_scaling}
\end{equation}
where $L$ is the characteristic length of the size, and $\tilde{f}_{M}$
is a universal function for this model. By comparing \eqref{eq:non_can_scaling}
and \eqref{eq:finite_size_scaling}, we observe that if we regard
the relative deviation of inverse temperature $\sigma\equiv\Delta\beta/\beta$
to be inversely proportional to the size of the system $L$, then
the magnetization of non-canonical states qualitatively obeys the
same scaling as that of finite-size systems.

\section{Correlation induced by non-canonical states}

\label{sec:non_can_correlation}

In this section, we discuss the correlation induced by the non-canonical
states. For a system composed of two components with the same energy
levels $E_{n}$, the total energy levels are given by $E_{(m,n)}=E_{m}+E_{n}$.
When the two-component system is in a canonical state with inverse
temperature $\beta$, the joint distribution is expressed as 
\[
P_{(m,n)}(\beta)=\frac{1}{Z_{2}}\mathrm{e}^{-\beta(E_{m}+E_{n})}=\frac{\mathrm{e}^{-\beta E_{m}}}{Z_{1}}\times\frac{\mathrm{e}^{-\beta E_{n}}}{Z_{1}}=P_{m}(\beta)P_{n}(\beta).
\]
Here, $Z_{2}=\sum_{m,n}\mathrm{e}^{-\beta(E_{m}+E_{n})}=Z_{1}^{2}$.
This means the joint distribution is a product of the distribution
for one particle, showing no correlation between components of the
canonical states.

For non-canonical states, the joint distribution becomes
\[
P_{(m,n)}^{(\mathrm{nc})}(\beta)=\int\mathrm{d}\beta'f(\beta',\beta)P_{(m,n)}(\beta)=\int\mathrm{d}\beta'f(\beta',\beta)P_{m}(\beta)P_{n}(\beta).
\]
While the distribution of a single component is still a non-canonical
distribution
\[
P_{n}^{(\mathrm{nc})}(\beta)=\sum_{m}P_{(m,n)}^{(\mathrm{nc})}(\beta)=\int\mathrm{d}\beta'f(\beta',\beta)P_{n}(\beta).
\]
Thus, one can show that for the non-canonical distribution, $P_{(m,n)}^{(\mathrm{nc})}(\beta)\neq P_{m}^{(\mathrm{nc})}(\beta)P_{n}^{(\mathrm{nc})}(\beta)$.
As a consequence, the information entropy shows no additivity.

\section{Emission Rate for Non-canonical Black Holes}

\label{sec:Emission-Rate-for}

The radiation field of a black body is described by a Boltzmann distribution,
where the particle emission rate for mode $\omega$ is:
\[
j=\frac{1}{4\pi^{2}}\frac{1}{\mathrm{e}^{\beta\omega}-1}.
\]
For a small deviation of inverse temperature, the non-canonical emission
rate is obtained by Eq. \eqref{eq:general_expansion} as 
\[
j^{(\mathrm{nc})}\approx\frac{1}{4\pi^{2}}\frac{1}{\mathrm{e}^{\beta\omega}-1}\left(1+\frac{1}{2}\Delta\beta^{2}\omega^{2}\mathrm{e}^{\beta\omega}\frac{1+\mathrm{e}^{\beta\omega}}{\left(1-\mathrm{e}^{\beta\omega}\right)^{2}}\right).
\]
With the condition $\exp(\beta\omega)\gg1$ and $\omega\Delta\beta\ll1$,
the non-canonical emission rate is reduced to 
\[
j^{(\mathrm{nc})}\approx\frac{1}{4\pi^{2}}\exp\left[-\beta\omega+\frac{1}{2}\omega^{2}\Delta\beta^{2}\right].
\]

Now we consider a small Schwarzschild black hole as a black body,
whose inverse temperature is distributed around $\beta=8\pi M$. With
the heat capacity for such black holes given by $C=-8\pi M^{2}$,
the deviation of the temperature is estimated as
\[
\Delta\beta^{2}\approx\left(\frac{\partial U}{\partial\beta}\right)^{-1}\left\langle \Delta E^{2}\right\rangle =\left|\frac{\beta^{2}}{C}\right|=8\pi.
\]
Hence, the non-canonical emission rate is obtained as follows:
\[
j^{(\mathrm{nc})}\approx\frac{1}{4\pi^{2}}\exp\left[-8\pi\omega\left(M-\omega/2\right)\right].
\]
This is precisely the result in Ref. \citep{Parikh2000}.

\bibliography{refs}